\documentclass[11pt]{article}
\usepackage{xcolor}
\usepackage{fullpage} 
\usepackage{authblk} 
\usepackage{url}
\usepackage[flushleft]{threeparttable}

\usepackage{caption}
\usepackage{amsmath,amssymb, amsfonts}
\usepackage{numprint}
\usepackage{float}
\usepackage{natbib}
\usepackage[title]{appendix}
\usepackage{graphicx} 
\usepackage{subcaption}
\graphicspath{ {./} }

\makeatletter
\def\thanks#1{\protected@xdef\@thanks{\@thanks
        \protect\footnotetext{#1}}}
\makeatother

\title{Recurrent Neural Networks for Multivariate Loss Reserving and Risk Capital Analysis}
\author{Pengfei Cai\textsuperscript{1}}
\author{Anas Abdallah\textsuperscript{1*}\thanks{Correspondence\textsuperscript{*}: Department of Mathematics and Statistics, McMaster University, 1280 Main Street West, Hamilton, Ontario, L8S 4K1, Canada. E-mail: anasabdallah@mcmaster.ca}}
\author{Pratheepa Jeganathan\textsuperscript{1*}}
\affil{Department of Mathematics \& Statistics, McMaster University}
\date{}
\begin{document}


\maketitle

\begin{abstract}

In the property and casualty (P\&C) insurance industry, reserves comprise most of a company’s liabilities. These reserves are the best estimates made by actuaries for future unpaid claims. Notably, reserves for different lines of business (LOBs) are related due to dependent events or claims. While the actuarial industry has developed both parametric and non-parametric methods for loss reserving, only a few tools have been developed to capture dependence between loss reserves. \textcolor{black}{This paper introduces the use of the Deep Triangle (DT), a recurrent neural network, for multivariate loss reserving, incorporating an asymmetric loss function to combine incremental paid losses of multiple LOBs. 
The input and output to the DT are the vectors of sequences of incremental paid losses that account for the pairwise and time dependence between and within LOBs.
 In addition, we extend generative adversarial networks (GANs) by transforming the two loss triangles into a tabular format and generating synthetic loss triangles to obtain the predictive distribution for reserves.
We call the combination of DT for multivariate loss reserving and GAN for risk capital analysis the extended Deep Triangle (EDT). To illustrate EDT, we apply and calibrate these methods using data from multiple companies from the National Association of Insurance Commissioners database. For validation, we compare EDT to the copula regression models and find that the EDT outperforms the copula regression models in predicting total loss reserve.} Furthermore, with the obtained predictive distribution for reserves, we show that risk capitals calculated from EDT are smaller than that of the copula regression models, suggesting a more considerable diversification benefit. Finally, these findings are also confirmed in a simulation study. \textcolor{black}{Our analysis demonstrates the potential of the EDT in predicting loss reserves and conducting risk capital analysis in practice.}

\end{abstract}


\section{Introduction}
\textcolor{black}{Insurance companies need to establish reserve funds to guarantee the future compensation of policyholders who have made claims.} To fulfill this obligation, insurers maintain claim reserves, ensuring the availability of funds for all necessary future payouts. These reserves are predominantly informed by historical claim data, which aids in estimating the expected future claims through various reserving methods. There are two primary approaches to loss reserving: the micro-level approach, which deals with individual claims, and the macro-level approach, which pertains to aggregated claims. 

The macro-level approach aggregates individual claims, organizing them into loss triangles according to accident and development years. For loss reserving, the chain ladder method has been widely used in practice with an assumption that claims will continue to develop similarly in the future. However, a notable limitation of this method is its exclusion of uncertainty in its calculations. \cite{mack_distribution-free_1993} present a method to compute the distribution-free standard error of the reserve based on the chain ladder method to address this gap. For further insight into stochastic loss reserving methods, specifically for a single line of business (LOB) and at the macro-level, the works of \cite{england_stochastic_2002} and \cite{wuthrich_stochastic_2008} provide comprehensive reviews.

Insurance companies typically assess risk measures across all their LOBs. These measures are calculated based on the predictive distribution of the total reserve, which includes reserves allocated for each LOB. Insurers must estimate this predictive distribution accurately, as it offers valuable insights for actuaries, especially in risk management.
When dealing with multiple LOBs, a common assumption is the independence of claims across different LOBs. In such scenarios, the portfolio's total reserve and risk measure is the sum of each LOB. This approach, known as the ``silo" method \citep{ajne1994additivity}, does not account for any diversification benefits. However, insurance companies often operate across multiple LOBs, where claims can be related. Therefore, it becomes imperative for insurers to account for the dependencies between claims in different LOBs. Acknowledging these dependencies is essential for accurately estimating total reserves and effectively leveraging diversification benefits in calculating risk capital. 

\textcolor{black}{The non-parametric and the parametric approach are the two primary approaches to modeling the dependence between two LOBs. In the non-parametric approach, the multivariate Mack model \citep{prohl2005multivariate} extends the traditional Mack model to capture dependence across multiple LOBs. The multivariate additive model \citep{ludwig2010calendar} uses flexible, data-driven methods to estimate dependence structures without assuming a specific functional form.
In the parametric approach, \cite{Shi2011} proposes a copula regression model for two LOBs, which links the claims with the same accident and development year with copulas.} This model assumes that claims from different triangles with the same accident year and development year are dependent, called pair-wise dependence. Moreover, studies have incorporated dependence between LOBs through Gaussian or Hierarchical Archimedean copulas and derived the predictive distribution of the reserve, the reserve ranges, and risk capital \citep{Abdallah2015,shi2012bayesian,de2012modeling}. \textcolor{black}{However, copula regression is limited in its flexibility regarding the marginal distribution and does not account for time dependence in the incremental paid losses. }

Various machine learning techniques have recently been developed in micro-level loss reserving for a single LOB. These methods are either tree-based learning methods or neural networks. The tree-based method is based on recursively splitting the claims into more homogeneous groups to predict the number of payments \citep{wuthrich2018machine}. It can include numerical and categorical attributes of the claimant, such as type of injury and payment history, as predictors. Moreover, \cite{Duval2019} uses a gradient boosting algorithm with a regression tree as the base learner for loss reserving.  \cite{gabrielli2018neural} propose separate over-dispersed Poisson models for claim counts and claim sizes embedded in neural network architecture. It allows learning the model structure using the boosting machines.

In the context of neural networks, \cite{Mulquiney2006artificial}  explored their use for predicting claim sizes, finding better performance compared to generalized linear models. \cite{wuthrich2018neural} extended Mack's Chain-Ladder method using neural networks for individual claim reserving, modeling development year ratios with claim features but without prediction uncertainty.  \cite{Taylor2019loss} noted that neural networks can capture interactions between covariates with minimal feature selection, though at the cost of interpretability and prediction accuracy.

 

 
\textcolor{black}{ 
Machine learning techniques are increasingly used in loss reserving; however, few models have been developed to capture dependence across LOBs using recurrent neural networks (RNN), as noted by \cite{cossette2021using}.
\cite{Kuo2019} introduced the Deep Triangle (DT) model for a single LOB, leveraging gated recurrent units (GRU) to model incremental paid losses and outstanding claims; its extension to multiple LOBs remains under-explored. 
In this paper, we extend the DT model to multivariate loss reserving by incorporating within and between LOBs dependencies, leveraging recent advancements in computing power and flexible frameworks for multi-LOB analysis. Our contributions include:
}

\begin{enumerate}
    \item \textcolor{black}{We propose an asymmetric loss function for DT, an unequal weighting scheme that combines multiple incremental paid losses when the volatilities in the paid losses between LOBs are different.  
    This enables DT to combine multiple loss functions using variance-based weighting to learn multiple objectives simultaneously.}

    \item \textcolor{black}{We introduce two techniques to generate predictive distribution for loss reserves. One approach is Generative Adversarial Networks (GANs), specifically conditional tabular GAN (CTGANs) and CopulaGANs, creating synthetic loss triangles \citep{goodfellow2014generative, patki2016synthetic, cote2020synthesizing}; the other is block bootstrapping training inputs and outputs, creating bootstrap samples. By integrating these approaches with DT, we propose three models: DT-CTGAN, DT-CopulaGAN, and DT-bootstrap, collectively referred to as Extended Deep Triangle (EDT). }

    \item \textcolor{black}{We investigate the optimal input sequence length for DT. Since accident years have varying lengths of development years, we examine the effect of different input sequence lengths and find that longer sequences generally yield improved performance.  }

    \item 
\textcolor{black}{We speed up the training in DT for the DT-GAN and DT-bootstrap. To reduce training costs, we pre-train DT on observed incremental paid losses and fine-tune the model weights on synthetic incremental paid losses, significantly improving computational efficiency.}

\end{enumerate}

The paper is structured as follows: Section 2 details the methodologies used for loss reserving and predictive distribution estimation, focusing on the proposed Extended Deep Triangle (EDT) approach. Section 3 applies and calibrates EDT using a dataset comprising personal and commercial automobile LOBs from multiple companies. We also compare the estimated risk capital across copula regression models, showing that EDT consistently yields lower risk capital estimates. Section 4 presents a simulation study that further demonstrates EDT’s ability to generate lower risk capital than copula regression models while achieving superior predictive performance. Finally, Section 5 summarizes our findings and discusses their implications.

\section{Methods}
\label{section:methods}

In this section, we introduce the notation, define the risk measures, describe the DT model with the asymmetric loss function, outline the preparation of training samples for CTGAN, CopulaGAN, and block bootstrapping, and detail the process for generating predictive distributions for loss reserves.
 
Let $X_{ij}$ denote the incremental paid losses of all claims in accident year $i$ ($1\le i \le I$) and development year $j$ ($1\le j \le I$). The accident year refers to the year the insured event happened. The first accident year is denoted with 1, and the most recent accident year is denoted with $I$. The development year indicates the time the payment is made. The incremental paid loss refers to all payments in development year $j$ for the claims in year $i$. For one company and one business line, the observed data $X_{ij}$ for $i=1,2,...,I$ and $j=1,2,...,I-i+1$ is shown in the upper triangle of Table \ref{table:loss_triangle}.



\begin{table}[h!]
\centering
\begin{threeparttable}
\caption{The Loss Triangle}
\label{table:loss_triangle}
\begin{tabular}{|c|c|c|c|c|c|}
\hline 
 & \multicolumn{5}{c|}{Development year j} \\ 
\hline 
Accident year i & 1 & 2 & ... & I-1 & I \\ 
\hline 
1 &  $X_{11}$ & $X_{12}$ & ... & $X_{1,I-1}$&$X_{1,I}$  \\ 
\hline 
2  & $X_{21}$ & $X_{22}$ & ... & $X_{2,I-1}$ &  \\ 
\hline 
... & ... & ... & ... &  &    \\ 
\hline 
I-1 & $X_{I-1,1}$ & $X_{I-1,2}$ &  &  &    \\ 
\hline 
I & $X_{I,1}$ &  &  &  &    \\ 
\hline 
\end{tabular} 

\begin{tablenotes}
      \small
      \item Note: The upper triangle is the loss triangle. The rows are accident year, and the columns are development year. $X_{ij}$ denotes the incremental paid loss in accident year $i$ and development year $j$.
    \end{tablenotes}
\end{threeparttable}

\end{table}

The incremental paid loss $X_{ij}$ is adjusted for each LOB's exposure to ensure comparability across accident years. The exposure variable, such as premiums or the number of policies, provides a scaling factor. The standardized incremental paid loss is then defined as  $Y_{ij}=X_{ij}/\omega_{i}$, where $\omega_{i}$ represents the exposure for the $i^\text{th}$ accident year. In the case of multiple LOBs from one company, the standardized incremental paid loss for the  $\ell^{\text{th}}$ LOB is denoted by $Y_{i,j}^{(l)}$, with its predicted value represented as $\hat{Y}_{ij}^{(\ell)}$.

To estimate the lower triangle values $X_{i j}^{(\ell)}$, we multiply  $\hat{Y}_{ij}^{(\ell)}$ by the corresponding exposure $\omega_{i}^{(\ell)}$.
This yields a point estimate of the outstanding claims for each LOB, given by $R^{(\ell)} = \sum_{i=2}^{I} \sum_{j=I-i+2}^{I} \omega_i^{(\ell)} \hat{Y}_{ij}^{(\ell)}$. Finally, the total reserve for the entire insurance portfolio is $R = \sum_{\ell=1}^{2} R^{(\ell)}$. 



In actuarial practice, reserve estimation extends beyond point estimates to include measures of reserve variability. Given the predictive distribution of reserves, denoted by $F_R$, we compute commonly used actuarial risk measures, such as value at risk (VaR) and tail value at risk (TVaR).

The $\text{VaR}_k$ is the $100*k$ percentile of R, i.e., $\text{VaR}_k(R)=F^{-1}_R(k)$ while $\text{TVaR}_k$ is the expected loss conditional on exceeding the $\text{VaR}_k$, i.e., $\text{TVaR}_k(R)=\mathbb{E}[R|R\ge \text{VaR}_k(R)]$. 

Tail Value-at-Risk (TVaR) is more informative than Value-at-Risk (VaR) in risk assessment. TVaR, a coherent risk measure captures the expected shortfall and adherence to the sub-additive property \citep{acerbi2002coherence}. This means that for any two reserves $R_1$ and $R_2$, corresponding to two LOBs, the combined risk measure $\rho$ of their sum is less than or equal to the sum of their individual risk measures. That is, $\rho(R_1+R_2) \le \rho(R_1) + \rho(R_2)$. This ensures that the total risk measure does not exceed the sum of the individual risk measures, reflecting diversification benefits in risk assessment. Contrarily, VaR lacks this sub-additivity, especially in skewed distributions, making TVaR a more reliable indicator in risk management. \textcolor{black}{From the insurance perspective, risk measures lacking the sub-additivity can be misleading because they can increase the company's liability, resulting in a larger tax deduction.}
 
From the TVaR, we calculate the risk capital, which is the difference between the risk measure and the liability value.
(see, e.g., \cite{dhaene2006risk}). 
Risk capital is also set aside as a buffer against potential losses from extreme events.  
In practice, the risk measure is set at a high-risk tolerance $k$, and the liability value is set at a lower risk tolerance between 60\% and 80\%, according to the risk appetite.

We define risk capital associated with total reserve $R$ as in \eqref{eqn:risk_capital}.
\begin{equation}
\text{Risk capital (R)} = \text{TVaR}_{k}(R) - \text{TVaR}_{60\%}(R).
\label{eqn:risk_capital}
\end{equation}


Exploring the diversification benefits between two LOBs is essential in risk management. The ``silo" approach computes risk measures for each LOB independently and aggregates them, disregarding potential diversification benefits. In contrast, our study compares the risk capital estimates obtained using the proposed method with those from the ``silo" approach, as defined in \eqref{eqn:risk_capital_gain}. This comparison highlights the impact of recognizing interdependencies between LOBs in risk capital (RC) estimation. Further details on this approach can be found in \cite{abdallah2023rank}.
 
\begin{equation}
 \text{Gain}=\Big(\text{RC}_{\text{Silo}}(R)-\text{RC}_{\text{Copula}}(R)\Big)/\text{RC}_{\text{Silo}}(R).
\label{eqn:risk_capital_gain}    
\end{equation}


Next, we describe the Deep Triangle model, which uses GRU to capture the complex dependence between two LOBs.

\subsection{Deep Triangle Architecture for Multivariate Sequence Prediction}

 \textcolor{black}{DT is a multi-task network that achieves better performance by jointly predicting incremental paid losses and outstanding claims.} In this section, we extend the DT methodology to predict incremental paid losses across two LOBs. Our approach leverages multivariate sequences to enhance the accuracy and robustness of the predictions.

\begin{figure}[ht]
    \centering
    \includegraphics[scale=0.45]{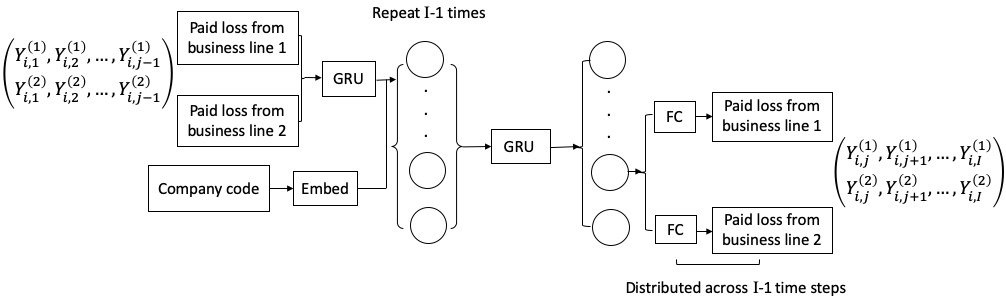}
    \caption{The DT architecture for multivariate sequence prediction. }
    \label{fig:dt_architecture}
\end{figure}

Figure \ref{fig:dt_architecture} illustrates the architecture of the DT model. We employ a vector sequence-to-sequence architecture to model the time series of incremental paid losses, effectively capturing both the pairwise dependence between two LOBs and the temporal dependence of incremental paid losses within each accident year \citep{sutskever_sequence_2014,srivastava2015unsupervised}. To our knowledge, this approach has not been previously explored in multivariate loss reserving analysis.
As depicted in Figure \ref{fig:dt_architecture}, consider $i^{\text{th}}$ accident year and $j^{\text{th}}$ development year. The input sequences are vectors of two elements: $\left(Y_{i,1}^{(1)}, Y_{i,2}^{(1)}, \ldots, Y_{i,j-1}^{(1)}\right)$ and $\left(Y_{i,1}^{(2)}, Y_{i,2}^{(2)}, \ldots, Y_{i,j-1}^{(2)}\right)$. The corresponding output sequences are vectors of two elements:
$\left(Y_{i,j}^{(1)}, Y_{i,j+1}^{(1)}, \ldots, Y_{i,I}^{(1)}\right)$ and $\left(Y_{i,j}^{(2)}, Y_{i,j+1}^{(2)}, \ldots, Y_{i,I}^{(2)}\right)$. We predict $I-j+1$ time steps into the future for the $j^{\text{th}}$ development year, resulting in an output sequence length of $I-1$.
Note that we assume the standardized incremental paid loss is independent across accident years. Since there is only one value for the last accident year, we do not use that incremental paid loss for training.

DT uses GRU to handle the time series of incremental paid losses for each accident year $i$ over development year $j$.  GRUs are preferred over Long Short-Term Memory (LSTM) networks due to their fewer training parameters and faster execution \citep{goodfellow2016deep}. The GRU processes each element in the input sequence vector and includes mechanisms to determine when a hidden state should be updated or reset at each time step. For each input sequence and for the current time step $\nu;\nu=1, \ldots , j-1$, the input to the GRU is $q_\nu= (Y_{i\nu}^{(1)}, Y_{i\nu}^{(2)})$ along with the previous time step's hidden state $h_{\nu-1}=(h_{\nu-1}^{(1)}, h_{\nu-1}^{(2)})$. The GRU outputs reset and update gates, $r_\nu$ and $z_\nu$ respectively, which take values between 0 and 1.

The reset gate $r_\nu$ and update gate $z_\nu$ for time step $\nu$ are computed as follows:
\begin{equation}
r_\nu = \sigma (W_{re} [h_{\nu-1}, q_\nu] + b_{re}),
\label{eqn:gru_reset_gate}
\end{equation}
and
\begin{equation}
z_\nu = \sigma (W_z [h_{\nu-1}, q_\nu] + b_z),
\label{eqn:gru_update_gate}
\end{equation}
where $W_{re}$ and $W_z$ are weight parameters, $b_{re}$ and $b_z$ are biases and $h_\nu$ is the hidden state value at $\nu$.
The weights and bias parameters are learned during training.
In \eqref{eqn:gru_reset_gate} and \eqref{eqn:gru_update_gate}, the sigmoid function $\sigma\left(.\right)$ is used to transform input values to the interval $(0,1)$. The candidate's hidden state at  $\nu$ is of the form

\begin{equation}
\tilde{h}_\nu = \text{tanh} (W_{hi} [r_\nu h_{\nu-1}, q_\nu] + b_{hi}). 
\end{equation}

The update gate $z_\nu$ determines the extent to which the new state $h_\nu$ is just the old state $h_{\nu-1}$ and how much of the new candidate state $\tilde{h}_\nu$ is used. The final update equation for the GRU is as follows:

\begin{equation}
h_\nu = z_\nu \tilde{h}_\nu + (1-z_\nu) h_{\nu-1}.
\label{eqn:gru_update}
\end{equation}

When the update gate $z_\nu$ is close to 0, the information from $q_\nu$ is ignored, skipping time step $\nu$ in the dependency chain. However, when $z_\nu$ is close to 1, the new state $h_\nu$ approaches the candidate state $\tilde{h}_\nu$. These designs help better capture sequence dependencies for $\left(Y_{i,1}^{(1)}, Y_{i,2}^{(1)}, \ldots, Y_{i,j-1}^{(1)}\right)$ and $\left(Y_{i,1}^{(2)}, Y_{i,2}^{(2)}, \ldots, Y_{i,j-1}^{(2)}\right)$. The outputs of the decoder GRU are then passed to two sub-networks of fully connected layers, which correspond to LOB 1 and LOB 2. Each consists of a hidden layer of 64 units, followed by an output layer of 1 unit representing the incremental paid loss at a time step $\nu$.  The final output sequences are denoted by ($\hat{Y}_{i,j}^{(1)}$, $\hat{Y}_{i,j+1}^{(1)}$, \ldots, $\hat{Y}_{i,I}^{(1)}$) and ($\hat{Y}_{i,j}^{(2)}$, $\hat{Y}_{i,j+1}^{(2)}$, \ldots, $\hat{Y}_{i,I}^{(2)}$).

\textcolor{black}{To enhance the robustness and generalizability of our model, we utilize data from multiple companies to train the DT model. We use $c_{ij}$ to denote the company code associated with $Y_{ij}^{(\ell)}$, which is processed through an embedding layer.}
This layer converts each company code into a fixed-length vector, where the length is a predetermined hyperparameter. In our implementation, we set the length as $C-1$, the number of companies minus one. This embedding process is an integral component of the neural network and is trained with the network itself rather than as a separate pre-processing step. Consequently, companies with similar characteristics are mapped to vectors that exhibit proximity regarding Euclidean distance.

\subsection{Training/Testing Setup}\label{train_test_setup}

The input for the training sample associated with accident year i ($1\le i \le I-1 $) and development 
year j ($2 \le j \le I+1-i $) are the sequences (mask,\ldots, mask, $Y_{i,1}^{(1)}$, $Y_{i,2}^{(1)}$, \ldots, $Y_{i,j-1}^{(1)}$) and (mask,\ldots, mask, $Y_{i,1}^{(2)}$, $Y_{i,2}^{(2)}$, \ldots, $Y_{i,j-1}^{(2)}$). The assumption is that $Y_{i,j}^{(1)}$ and $Y_{i,j}^{(2)}$ are predicted using the past $I-1$ time steps.   Note that there is no historical data before development year 1.  Thus, we use a mask value where $j < 1$ and $j > I$. \textcolor{black}{ Masking selectively ignores certain parts of the sequences during training. If the value at a timestep is equal to the mask value, that timestep is skipped in subsequent calculations, including the computation of the loss for backpropagation.}

The output for the training sample associated with accident year i ($1\le i \le I-1 $) and development 
year j ($2 \le j \le I+1-i $) are the sequences ($Y_{i,j}^{(1)}$, $Y_{i,j+1}^{(1)}$, \ldots, $Y_{i,I+1-i}^{(1)}$, mask, \ldots, mask) and ($Y_{i,j}^{(2)}$, $Y_{i,j+1}^{(2)}$, \ldots, $Y_{i,I+1-i}^{(2)}$, mask, \ldots, mask). Note that the output sequences also consist of $I-1$ time steps. We use a mask value because we do not have the lower part of the triangle.

The training data is randomly split into training and validation sets using an 80-20 split. When splitting, the training data corresponding to the same accident year and development year from different companies stay in the same training or validation sets.  We train the DT model for a maximum of 1000 epochs, employing an early stopping scheme. If the loss on the validation set does not improve over a 100-epoch window, we stop training and keep the weights on the epoch with the lowest validation loss. In the DT, we initialize the neural networks with random weights using the He initialization technique \citep{he2015delving}, recommended for ReLU activation function \citep{murphy2022probabilistic}. 

For the DT model, the loss function is the average over the predicted time steps of the mean squared error of predictions. For each output sequence ($\hat{Y}_{i,j}^{(1)}$, $\hat{Y}_{i,j+1}^{(1)}$, \ldots, $\hat{Y}_{i,I+1-i}^{(1)}$, mask, \ldots , mask) and ($\hat{Y}_{i,j}^{(2)}$, $\hat{Y}_{i,j+1}^{(2)}$, \ldots, $\hat{Y}_{i,I+1-i}^{(2)}$, mask, \ldots, mask), the symmetric loss is defined as

\begin{equation}
\frac{1}{I-i+1-(j-1)} \sum_{\nu=j}^{I+1-i} 
\frac{(\hat{Y}^{(1)}_{i,\nu}-Y^{(1)}_{i,\nu})^2+(\hat{Y}^{(2)}_{i,\nu}-Y^{(2)}_{i,\nu})^2}{2}.
\label{eqn:symmetric_loss}
\end{equation}

\textcolor{black}{We define an asymmetric loss function as}

\begin{equation}
\frac{1}{I-i+1-(j-1)} \sum_{\nu=j}^{I+1-i} 
\frac{1}{2{(\sigma_{i,j}^{(1)})}^2}{(\hat{Y}^{(1)}_{i,\nu}-Y^{(1)}_{i,\nu})^2+\frac{1}{2{(\sigma_{i,j}^{(2)})}^2}(\hat{Y}^{(2)}_{i,\nu}-Y^{(2)}_{i,\nu})^2},
\label{eqn:asymmetric_loss}
\end{equation}
\textcolor{black}{where ${(\sigma_{i,j}^{(1)})}^2$ and ${(\sigma_{i,j}^{(2)})}^2$ are variances for sequences ($Y_{i,j}^{(1)}$, $Y_{i,j+1}^{(1)}$, \ldots, $Y_{i,I+1-i}^{(1)}$, mask, \ldots, mask) and ($Y_{i,j}^{(2)}$, $Y_{i,j+1}^{(2)}$, \ldots, $Y_{i,I+1-i}^{(2)}$, mask, \ldots, mask), respectively. The volatilities in the paid losses are different between the two LOBs, and we use uncertainty-based weighting to balance the two prediction tasks. When calculating the variances ${(\sigma_{i,j}^{(1)})}^2$ and ${(\sigma_{i,j}^{(2)})}^2$ , we exclude the mask value from the sequences. }

To optimize the parameters for the DT model, we employ the AMSGRAD variant of the Adaptive Moment Estimation (ADAM) algorithm \citep{Reddi2018, kingma2014adam}. AMSGRAD is chosen specifically due to its ability to manage the high variability in gradients that arise from the small size of the training sample, a common issue in stochastic gradient descent (SGD) methods. AMSGRAD addresses this by incorporating the gradients' moment into the parameter update process, thus offering a more stable and effective optimization in scenarios with limited data.

Next, we predict future incremental paid loss with the trained and validated DT and obtain a point estimate of the reserve. The input for the testing sample associated with accident year i ($2\le i \le I $) and development 
year j ($j=I+2-i$) are the sequences (mask, \ldots, mask,$Y_{i,1}^{(1)}$, $Y_{i,2}^{(1)}$, \ldots, $Y_{i,I+1-i}^{(1)}$) and (mask, \ldots, mask, $Y_{i,1}^{(2)}$, $Y_{i,2}^{(2)}$, \ldots, $Y_{i,I+1-i}^{(2)}$). There are $I-1$ testing samples whose accident year and development year satisfy $i+j=I+2$ $(2\le i \le I )$.
For accident year 1, we have all the data from development year 1 to development year $I$.
The input sequences for testing also consist of $I-1$ time steps.  At each accident year and development year for which we have data, we predict future incremental paid loss  ($\hat{Y}_{i,I+2-i}^{(1)}$, $\hat{Y}_{i,I+3-i}^{(1)}$, \ldots, $\hat{Y}_{i,I}^{(1)}$) and ($\hat{Y}_{i,I+2-i}^{(2)}$, $\hat{Y}_{i,I+3-i}^{(2)}$, \ldots, $\hat{Y}_{i,I}^{(2)}$).
Next, we obtain a point estimate of the outstanding claims for each LOB by $R^{(\ell)} = \sum_{i=2}^{I} \sum_{j=I+2-i}^{I} \omega_i^{(\ell)} \hat{Y}_{ij}^{(\ell)}$.

Once total loss reserves are estimated using the DT model, our approach includes generating the predictive distribution of total loss reserves.

\subsection{Predictive Distribution of the Total Reserve}

We outline bootstrapping in the context of DT to generate the predictive reserve distribution, referred to as DT-bootstrap. Specifically, we use the block bootstrap technique \citep{buhlmann2002bootstraps} to resample training data because we identify exchangeability \textcolor{black}{between different pairs of input sequences} for DT described in Section \ref{train_test_setup} with block size $I$. First, we randomly split the training data into training and validation sets using an 80-20 split described in Section \ref{train_test_setup}. Suppose $I=10$. For each company, we have 36 training sequences and 9 validation sequences. Let $\boldsymbol{X}_n$ ($1\le n \le 36$) denote the training input sequences (mask,..., mask, $Y_{i,1}^{(1)}$, $Y_{i,2}^{(1)}$, ..., $Y_{i,j-1}^{(1)}$) and (mask ,..., mask, $Y_{i,1}^{(2)}$, $Y_{i,2}^{(2)}$, ..., $Y_{i,j-1}^{(2)}$) from one company. We also let $\boldsymbol{Y}_n$ ($1\le n\le 36$) denote the training output sequences ($Y_{i,j}^{(1)}$, $Y_{i,j+1}^{(1)}$, ..., $Y_{i,11-i}^{(1)}$, mask, ... , mask) and ($Y_{i,j}^{(2)}$, $Y_{i,j+1}^{(2)}$, ..., $Y_{i,11-i}^{(2)}$, mask, ... , mask). Our original training data are ($\boldsymbol{X}_1$,$\boldsymbol{Y}_1$) , ... , ($\boldsymbol{X}_{36}$,$\boldsymbol{Y}_{36}$). We draw bootstrap training data ($\boldsymbol{X}_1^*$,$\boldsymbol{Y}_1^*$) , ... , ($\boldsymbol{X}_{36}^*$,$\boldsymbol{Y}_{36}^*$) randomly with replacement from the original training set. The training data with the same accident year and development year of different companies stay together during bootstrapping. We apply the same procedure to the validation data.

\textcolor{black}{
Next, we adapt Generative Adversarial Nets (GANs), as introduced by \cite{goodfellow2014generative}, to generate synthetic loss triangles and to generate the predictive distribution of the total reserve. This approach provides a novel way of applying advanced machine learning methods to the traditional actuarial problem of reserve estimation. GAN generates new data based on learned distributions from the original data. Bootstrap resamples the data with replacement to create multiple simulated data. Bootstrap may not sufficiently capture the underlying data distribution, especially in complex scenarios. }

\begin{figure}[ht]
    \centering
    \includegraphics[scale=0.35]{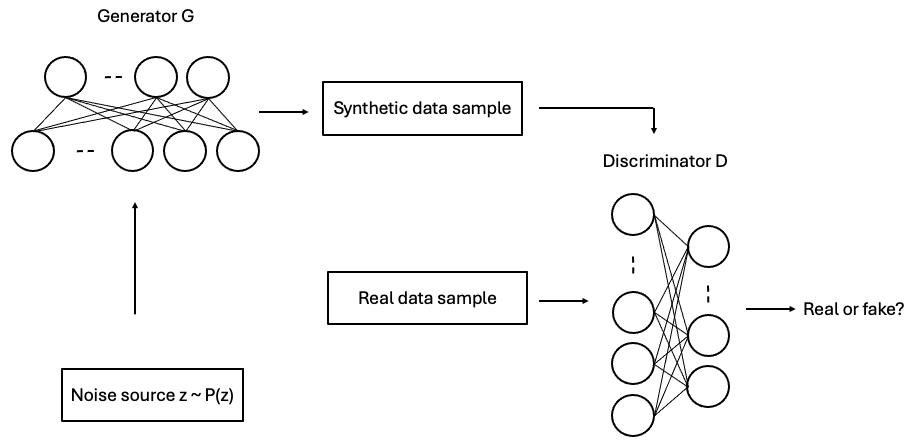}
    \caption{The architecture of GANs.}
    \label{fig:gan_architecture}
\end{figure}

GAN simultaneously trains two models: a generative model, G, which captures the data distribution and generates new data, and a discriminative model, D, which outputs the probability of how likely the generated data belongs to the training data. Figure \ref{fig:gan_architecture} shows the relationship between the generator G and the discriminator D. The generator G generates realistic samples while the discriminator D distinguishes between genuine and counterfeit samples. The generator G takes some noise $z$ as input and outputs a synthetic sample.

\textcolor{black}{
The traditional GAN model utilizes latent variable
z sampling from a standard multivariate normal distribution. However, for our purposes, we require a GAN approach that can capture the pairwise dependence between two LOBs and the sequential structure inherent in the standardized incremental paid losses. To address this, we extend the GAN method for tabular data generation, considering the pairwise dependency and the sequential structure.}

\textcolor{black}{
To generate synthetic data from loss triangles, we employ conditional tabular GAN (CTGAN), a GAN variant \citep{xu2019modeling}, which is adept at modeling dependencies in data. 
For CTGAN, numerical inputs $Y_{i j}^{(\ell)}$ are normalized to fit within the (-1, 1) range using mode-specific normalization. Each $Y_{i j}^{(\ell)}$ is represented as a one-hot vector $\beta_{i j}^{(\ell)}$, indicating the mode, and a scalar $\alpha_{ij}^{(\ell)}$, indicating the value within the mode.  The company code categorical variable $c_{ij}$ is represented as a one-hot vector $d_{ij}$. }

\textcolor{black}{
CTGAN generates data conditioned on additional information by combining random noise $z$ sampling from a standard normal vector with a condition (such as company code $c_{ij}$). This is done by concatenating the noise $z$ and the condition and then passing the combined input through the generator network $G$, which learns to generate synthetic data that satisfies the given condition. 
Additionally, to maintain the sequential integrity of the data, we generate synthetic data for each development year separately. This approach ensures that the sequential properties of standardized incremental paid losses are preserved, allowing for more accurate and realistic simulation of loss triangles across multiple LOBs.}

\textcolor{black}{
Generating a synthetic loss triangle involves a three-stage process:}
\textcolor{black}{
(1) Combination of Data: For each development year $j$ ($1\le j \le I$), we combine $Y_{i j}^{(1)}$ and $Y_{i j}^{(2)}$ of all accident years $i$ ($1\le i \le I$) from all companies into one table.  The first column of the table is $Y_{i j}^{(1)}$ of all accident years from the personal auto line. The predicted loss from the DT model is used if the $Y_{i j}^{(1)}$ is not available. Similarly, the data from the commercial auto line is used for the second column. The third column of the table is the company code.}

\textcolor{black}{
(2) GAN Model Training: We train a GAN model for each development year $j$ using the combined data from (1). The representation $r_{ij}$ of a row in the combined table is the concatenation of the three columns: $r_{ij}=\alpha_{ij}^{(1)} \oplus \beta_{ij}^{(1)} \oplus \alpha_{ij}^{(2)} \oplus \beta_{ij}^{(2)} \oplus d_{ij}$.  This training enables the GAN model to learn the underlying distribution of the combined data in (1).}

\textcolor{black}{
(3) Sampling and Loss Triangle Formation: After training, we use the GAN to sample $I$ new rows for each development year $j$ for each company. These sampled data are then sequentially arranged according to their development years. We remove the lower triangle to form a new loss triangle, ensuring the structure aligns with the loss triangle format.}

\textcolor{black}{
Note that the loss triangles from the two LOBs are on different scales, presenting a challenge for effective modeling. To address this, we employ a CopulaGAN \citep{patki2016synthetic}, which leverages the scale-invariant property of copulas to define the covariance of $z$. In addition to the marginal distributions, CopulaGAN uses a Gaussian copula. CopulaGAN is a variation of the CTGAN model that leverages the CDF-based transformation applied by Gaussian Copula, making it easier for the underlying CTGAN model to learn the data. For computing,  we use the SDV \citep{patki2016synthetic} library to build GAN models by constructing the input as follows.  }

\textcolor{black}{
\begin{itemize}
    \item[$(1)$] Let the marginal CDFs of columns $Y_{i j}^{(1)}$ and $Y_{i j}^{(2)}$  be as $F_1$ and $F_2$, respectively.
    \item[$(2)$] Go through the table row-by-row. Each row is denoted as $y=\left(Y_{i j}^{(1)}, Y_{i j}^{(2)}\right)$.
    \item[$(3)$] Transform each row using the inverse probability transform: 
    \begin{equation*}
    z=\left[\Phi^{-1}\left(F_1\left(Y_{i j}^{(1)}\right)\right), \Phi^{-1}\left(F_2\left(Y_{i j}^{(2)}\right)\right)\right]
    \end{equation*} where $\Phi^{-1}\left(\cdot\right)$ is the inverse CDF of the Gaussian distribution. 
    \item[$4)$] After all the rows are transformed, estimate the covariance matrix $\Sigma$ of the transformed values.
\end{itemize}
}

\textcolor{black}{
The parameters for each column distribution and the covariance matrix $\Sigma$ are used in the generative model for that table. The CDF transformed data $F_1(Y_{i j}^{(1)})$ and $F_1(Y_{i j}^{(1)})$ are then fed into the CTGAN architecture. After generating synthetic data in the transformed space, CopulaGAN uses the inverse CDF to bring the data back to the original space. For the new table generated using CopulaGAN, we consider only the upper loss triangle a new sample.}

Finally, we apply the DT model to the loss triangles generated by the previously described procedures (GAN or bootstrapping). Repeating this procedure many times enables us to construct a predictive distribution for the reserve. Subsequently, the risk capital gain is calculated using the methodology outlined in \eqref{eqn:risk_capital_gain}. The proposed EDT method, which includes DT-GAN and DT-bootstrap, integrates the predictive capabilities of the DT model with the distributional insights provided by GAN or bootstrapping, offering a comprehensive view of the reserve estimation and its associated risks.

\section{Applications}\label{application}
\subsection{Data Description}

In this section, we demonstrate the proposed EDT approach on a real dataset. 
We use 30 pairs of 30 companies loss triangles from Schedule P of the National Association of Insurance Commissioners (NAIC) database \citep{meyers2011loss} to illustrate and compare EDT. Each pair comprises two loss triangles from personal and commercial auto LOBs and is associated with a company code. Each triangle contains incremental paid losses for accident years 1988-1997 and ten development years. 
Here, we demonstrate a prediction and risk capital analysis for a major US property-casualty insurer.
Appendix Table \ref{tab:data_shi_lob1} and Appendix Table \ref{tab:data_shi_lob2} show the incremental paid losses for this company's personal and commercial auto LOBs. 

\subsection{Prediction of Total Reserve}

First, we apply the DT model to the claims in the two LOBs from 30 companies. In the pairwise training sample, the first component is the incremental paid loss from the personal LOB, and the second component corresponds to the incremental paid loss from the commercial LOB. We use the same standardization to get $Y_{i,j}$ as in Section \ref{section:methods}.

To train the DT, we consider the incremental paid losses up to 1997, which is the current calendar year for this dataset. \textcolor{black}{We predict the reserve using DT with both the symmetric loss function as in \eqref{eqn:symmetric_loss} and the asymmetric loss function as in \eqref{eqn:asymmetric_loss}. Table \ref{tab:independence_copula} displays the predicted reserves from DT alongside the actual reserves. In terms of the percentage error from the actual reserve, DT with the asymmetric loss function generates a more accurate estimation of the reserve, which is shown in Table \ref{tab:reserve_comparison_real}. Moreover, bias is greatly reduced for the most volatile commercial LOB's reserve when introducing the asymmetric loss function. 
} 

Next, we apply the copula regression model to the two loss triangles from the major US property-casualty insurer. \textcolor{black}{ We refer to the details of the copula regression in Appendix \ref{section:copula_regression}}. For the marginal distribution, we use the log-normal
and the gamma distributions for personal and commercial LOBs \citep{Shi2011}, respectively. 
We consider the systematic component $\eta_{ij}=\mu_{ij}$ for the log-normal distribution with location parameter $\mu_{ij}$ and shape parameter $\sigma$. For the gamma distribution with location parameter $\mu_{ij}$ and shape parameter $\phi$, we use $\eta_{ij}=\log(\mu_{ij}\phi)$.

To model the dependence between the two LOBs,
we use the Gaussian copula, Frank copula, and Student's t copula. These copula functions are specified using the R package \texttt{copula} \citep{hofert2020copula}.
The \texttt{gjrm} function from the R package \texttt{GJRM} estimates the copula regression model \citep{Marra2023}. The log-likelihood, Akaike information criterion (AIC), and Bayesian information criterion (BIC) for all copula models are provided in Appendix Table \ref{tab:copula_fit_statistics}. The Student's t copula leads to the smallest AIC and BIC adjusted for the unseen lower triangle. According to Table \ref{tab:independence_copula}, independence and copula models generate comparable point estimates for the total reserve, about 7 million dollars.

\begin{table}[h!]
\centering
\caption{Point estimates of the reserves from DT and copula regression models.}
\begin{threeparttable}
\begin{tabular}{c c c c}
\hline 
 & \multicolumn{3}{c}{Reserves} \\ 
\cline{2-4}
Model & {\hspace{0.1in}LoB 1, $R_1$} & {LoB 2, $R_2$} & {Total, $R$} \\ 
\hline 
DT (symmetric)& 7 756 417 & 327 517 & 8 083 934 \\
\textcolor{black}{DT (asymmetric)} & 7 781 299 & 324 024 & 8 105 323 \\
Product Copula & 6 464 083 & 490 653 & 6 954 736 \\ 
Gaussian Copula & 6 423 246 & 495 925 & 6 919 171 \\
Frank Copula & 6 511 360 & 487 893 & 6 999 253 \\
Student's t Copula & 6 800 554 & 554 426 & 7 354 980 \\
Actual Reserve & 8 086 094 & 318 380 & 8 404 474 \\
\hline 
\end{tabular} 
\end{threeparttable}
\label{tab:independence_copula}
\end{table}


\textcolor{black}{In addition to DT, Table 3 shows the percentage error of prediction to the actual reserve for the coupla regression model.  Interestingly, the DT model provides a more accurate point estimation of the reserve for personal and commercial auto LOBs. This improved performance is attributed to the neural network's ability to learn complex non-linear relationships of incremental paid losses between LOBs and within accident years \citep{murphy2022probabilistic}. }

\begin{table}[h!]
\centering
\scalebox{0.85}{
\begin{threeparttable}
\caption{Performance comparison using percentage error of actual and estimated loss reserve.}
\label{tab:reserve_comparison_real}

\begin{tabular}{l c c c c c}
\hline
 LOB & DT (symmetric)& DT (asymmetric)&Product Copula &   Gaussian Copula &   Frank Copula  \\  
 \hline
 Personal Auto  & {-4.1\%} & \textbf{-3.8\%}&{-20.1\%} & {-20.6\%} & {-19.5\%} \\ 
 Commercial Auto   & {2.9\%} & \textbf{1.8\%} & {54.1\%} & {55.8\%} & {53.2\%} \\ 
 Total  & {-3.8\%} & \textbf{-3.6\%} & {-17.3\%} & {-17.7\%} & {-16.7\%}\\ 
 \hline
\end{tabular}

\begin{tablenotes}
      \small
      \item Note: The best metric for each LOB is in bold.
    \end{tablenotes}
\end{threeparttable}
}
\end{table}
 
\textcolor{black}{For a fair comparison between copula regression and DT, we also utilize 30 companies' data for copula regression. The results in Appendix \ref{app:copula_regression30} show similar results to those with one company. However, the company heterogeneity should be modeled as a random effect, and to our knowledge, there is no software implementation of such a model to be used in this analysis.}

\subsection{Predictive Distribution of Total Reserve}

\begin{figure}[h!]
    \centering
    \includegraphics[scale=0.4]{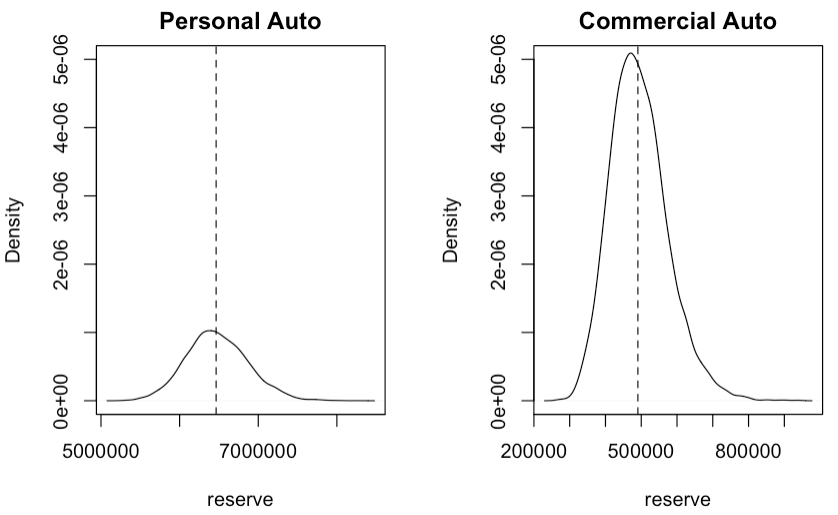}
    \caption{Marginal predictive distributions of the reserves from parametric bootstrapping. \\ {\small Note: The vertical dotted lines indicate the maximum likelihood estimates of the reserve.}}
    \label{fig:predictive_distribution_2lines}
\end{figure}

 First, we obtain the marginal predictive distribution of reserves to evaluate the diversification benefits. We use parametric bootstrapping for log-normal and gamma models to approximate the marginal predictive distribution for personal and commercial LOBs as illustrated in Figure \ref{fig:predictive_distribution_2lines}. Our results show different predictive distributions for the reserves in the two LOBs; notably, personal LOBs exhibit heavier tails than commercial LOBs. \textcolor{black}{ This differentiation enables us to apply different values to business volumes, taking into account the dependence between LOBs.} For instance, increasing the volume of the commercial LOB can enhance the advantages of diversification. Next, we generate the predictive distributions of the total reserves.

 To obtain the predictive distribution of the reserve for the DT model, we first utilize the CTGAN model. We train the CTGAN model with data from 30 companies for each development year. New incremental paid losses are generated for each of the ten development years. The newly generated data for each development year are then stacked in the order of development years to form new upper loss triangles. The DT model estimates reserves for these newly created loss triangles. This process is repeated multiple times to construct a predictive reserve distribution. In addition to CTGAN, CopulaGAN and block bootstrapping methods are employed to create new upper triangles to generate the predictive distribution of the reserve. 

\textcolor{black}{To reduce the computational expense associated with EDT, stemming from training numerous
DTs for GAN samples, we leverage the trained model
on observed data to fine-tune weights for new samples. For newly generated sample, DT takes two minutes to run when trained from a random weight initialization, and about one minute when trained from a saved model that was previously fitted on real data. It takes about 4 hours to obtain the predictive distribution using parallel computing with 32 CPUs. 
}

For copula regression models, as discussed in Appendix \ref{section:copula_regression}, we used the parametric bootstrap to generate the predictive distributions of the reserves.
Table \ref{tab:bootstrap_bias} shows the estimated reserve, bias, and standard errors for the different models. The percentage bias is computed as the percentage error between the bootstrap mean reserves and estimated loss reserves. It has been observed that the standard deviations from the DT-GAN are smaller than those from the copula regression models, and the biases except the DT-bootstrap are slightly higher, likely due to heterogeneity across companies and between LOBs. This is also corroborated in Table \ref{tab:reserve_comparison_real}, which shows positive and negative bias for the commercial and personal LOB, respectively. CTGAN uses Gaussian mixtures to model distributions when generating synthetic loss triangles \citep{xu2019modeling}. These modeling assumptions in GANs introduce a certain bias compared to the bootstrapping method used in DT. However, the DT-CopulaGAN model, by capturing the inter-LOB dependence, generates a smaller risk capital than the bootstrapping approach in DT, as discussed in the next subsection. The coefficient of variation (CV) can be used to measure the risks when the insurance company has more than one LOB.   Based on the CV in Table \ref{tab:bootstrap_bias}, all the copula regression models and EDT have CVs smaller than one, which comply with the insurance standards. \textcolor{black}{However, the EDT stands out because its CVs are consistently smaller than the copula regression models.}

\begin{table}[h!]
\centering
\caption{Bias, Standard deviation, Coefficient of variation (CV)}
\begin{tabular}{l c c c c c}
\hline 
 & {Reserve} & {Bootstrap reserve} & Bias & {Std. dev.} & {CV} \\ 
\hline 
DT-CTGAN & 8 105 323 & 8 261 718 &1.93\% &197 465 &0.024\\
DT-CopulaGAN &8 105 323 & 8 255 638 & 1.85\% & 196 791 & 0.023\\
DT-bootstrap & 8 105 323 & 8 137 107 & 0.39\% & 235 304 & 0.029\\
Product Copula & 6 954 736  & 6 972 792 & 0.26\%  & 399 758 & 0.057\\
Gaussian Copula & 6 919 171   & 6 941 806  & 0.33\%  & 368 555 &  0.053\\
Frank Copula & 6 999 253   & 7 043 309  & 0.63\% & 388 357 & 0.056 \\
Student's t Copula & 7 354 980 & 7 383 876 & 0.39\% & 402 127 & 0.054\\
\hline 
\end{tabular} 
\label{tab:bootstrap_bias}
\end{table}

 Both the bootstrap reserve and standard deviation corresponding to the Student's t copula are larger than those of other copula models due to possibly overfitting for the Student's t copula. In addition, the dependence captured by the Student's t copula regression is not significant, as seen from Appendix Table \ref{app_tab:theta_confidence_interval}. Moreover, Table \ref{tab:bootstrap_bias} shows the standard error from the Student's t copula regression is comparable to that of the product copula regression. Though the dependence parameter from Student's t copula is insignificant, we keep it for comparison purposes in risk capital analysis.

\begin{figure}[h!]
\begin{subfigure}{.5\textwidth}
    \centering
    \includegraphics[width=.94\linewidth]{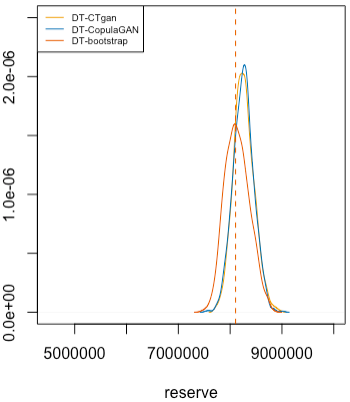}
\end{subfigure}%
\begin{subfigure}{.5\textwidth}
    \centering
    \includegraphics[width=.95\linewidth]{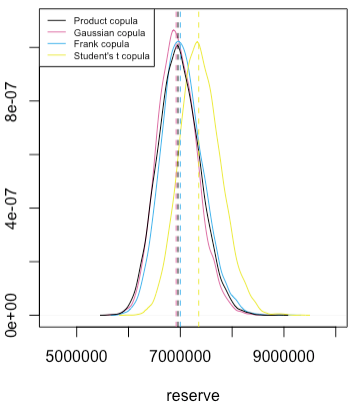}
\end{subfigure}
\caption{Predictive distributions of total reserves from EDT and copula regression. \\ {\small  Note: The vertical dotted lines indicate the estimated reserves for each model.}}
\label{fig:predictive_distribution_reserve}
\end{figure}

We compare the predictive distributions of the reserves for EDT and the copula regression in Figure \ref{fig:predictive_distribution_reserve}. This shows the bootstrap mean reserves of these models are pretty similar, except for the reserve from the model with Student's t copula. In summary, for this insurer, the dependence between triangles does not change much in the point estimate of reserves using copula regression models. Though the point estimates are similar, the dependencies between LOBs affect the reserve's predictive distribution, which helps diversify risk within the portfolio.

\begin{figure}[h!]
    \centering
    \includegraphics[scale=0.4]{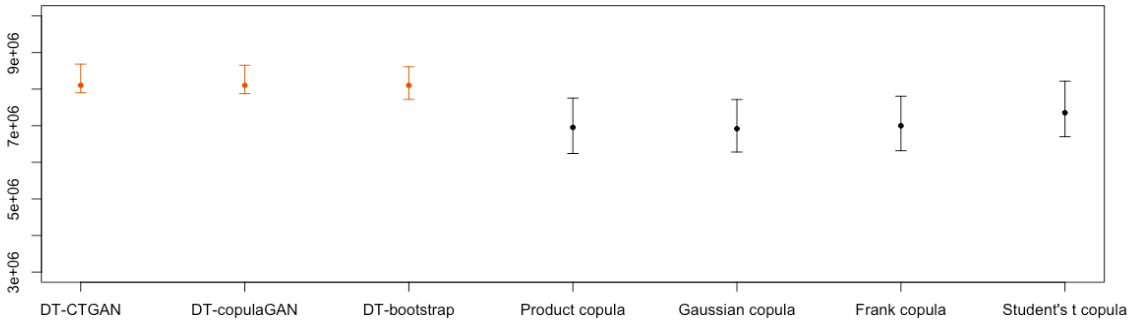}
    \caption{95\% confidence interval for the total reserves for different models. }
    \label{fig:conf_interval_real}
\end{figure}

Using the predictive distributions we generated, Table \ref{tab:conf_interval_real} and Figure \ref{fig:conf_interval_real} showcase the 95\% confidence interval of the total reserve, where the lower bound is the 2.5th percentile of the predictive distribution and the upper bound is the 97.5th percentile of the predictive distribution. We observe that the EDT models generate the narrowest confidence intervals.

\begin{table}[h!]
\centering
\caption{95\% confidence intervals for the total reserve using the predictive distribution.}
\begin{tabular}{l}
\hline 
   \\ 
\hline 
DT-CTGAN \\
DT-CopulaGAN \\
DT-bootstrap  \\ 
Product Copula  \\
Gaussian Copula \\
Frank Copula \\
Student's t Copula \\
\hline 
\end{tabular}%
\begin{tabular}{c c}
\hline
{Lower bound} &{Upper bound} \\
\hline
  7 900 272 & 8 683 653 \\ 
  7 875 828 & 8 653 414 \\ 
  7 719 487 & 8 613 324 \\ 
6 241 016 & 7 756 950 \\ 
  6 280 339 & 7 715 924 \\ 
  6 315 438 & 7 807 835 \\ 
  6 698 055 & 8 219 462 \\ 
\hline
\end{tabular}
\label{tab:conf_interval_real}
\end{table}

\subsection{Risk Capital Calculation}

\textcolor{black}{In actuarial practice, a common method to calculate the risk measure for the entire portfolio, which includes both personal auto and commercial auto LOBs, is the ``silo" method, as outlined in the methods section.} In our analysis, ``Silo-GLM" represents the aggregate of risk measures derived from copula regression models. In contrast, ``Silo-DT" refers to the sum of risk measures obtained from the EDT model.

To evaluate the benefits of diversification, we compared these silo methods with the risk measures calculated from both EDT and copula regression models. The comparative results are displayed in Table \ref{tab:bootstrap_risk_capital}. Our findings indicate that due to the negative association between the two LOBs (as detailed in Appendix Section \ref{app:kendall_tau}), the risk measures from both EDT and copula regression models are lower than their respective silo totals. This suggests that incorporating the interdependencies between LOBs into the risk measurement process can yield lower risk estimates, highlighting the potential value of these more integrated approaches in risk management.
 
\begin{table}[h!]
\centering
\caption{Risk capital estimation for different methods.}
\scalebox{0.85}{
\begin{tabular}{c c c c c c c}
\hline
\textbf{Risk measure} & {TVaR (60\%)} & {TVaR (80\%)} & {TVaR (85\%)} &{TVaR (90\%)} &{TVaR (95\%)} & {TVaR(99\%)}\\ 
\hline 
Silo-DT &  8 105 434 & 8 217 523 & 8 258 835 & 8 317 629 & 8 399 375 & 8 558 981 \\
DT-CTGAN & 8 452 870 & 8 549 295 & 8 582 802 & 8 627 247 & 8 699 935 & 8 846 865 \\
DT-CopulaGAN & 8 443 684 & 8 532 939 & 8 564 483 & 8 606 020 & 8 666 096 & 8 789 649\\
DT-Bootstrap & 8 370 792 & 8 480 870 & 8 519 400 & 8 565 251 & 8 632 670 & 8 743 527 \\ 
Silo-GLM & 7 442 692 & 7 671 633 & 7 756 992 & 7 872 138 & 8 060 489 & 8 460 435 \\ 
Product copula & 7 367 695 & 7 553 768 & 7 621 203 & 7 710 435 & 7 847 773 & 8 126 433 \\
Gaussian copula & 7 313 951 & 7 490 387 & 7 556 029 & 7 644 886 & 7 782 646 & 8 054 737 \\
Frank copula & 7 424 807 & 7 616 405 & 7 685 514 & 7 776 754 & 7 921 574 & 8 202 695  \\ 
Student's t copula & 7 779 447 & 7 969 264 & 8 038 676 & 8 129 761 & 8 275 013 & 8 563 869  \\
\hline
\hline
\textbf{Risk capital} & & & & & \\
\hline
Silo-DT & & \textbf{112 089} & \textbf{153 401} & \textbf{212 195} & \textbf{293 941} & \textbf{453 547} \\
DT-CTGAN & &96 425 & 129 932 & 174 377 & 247 065 & 393 995\\
DT-CopulaGAN & &89 255 & 120 799 & 162 336 & 222 412 & 345 965 \\
DT-Bootstrap & &110 078 & 148 608 & 194 459 & 261 878 & 372 735 \\
Silo-GLM &        & 228 941 & 314 300 & 429 446 & 617 797 & 1 017 743 \\
Product copula &      & 186 073 & 253 508 & 342 740 & 480 078 & 758 738 \\ 
Gaussian copula &        & 176 436 & 242 078 & 330 935 & 468 695 & 740 786 \\ 
Frank copula &        & 191 598 & 260 707 & 351 947 & 496 767 & 777 888 \\
Student's t copula & & 189 817 & 259 229 & 350 314 & 495 566 & 784 422 \\
\hline 

\end{tabular} 
}
\label{tab:bootstrap_risk_capital}
\end{table}

Next, we calculate risk capital as defined in (\ref{eqn:risk_capital}) from the predictive distribution of the reserves to set up a buffer from extreme losses.
Table \ref{tab:bootstrap_risk_capital} shows the risk capital required under the Silo-DT method is less than that of Silo-GLM, suggesting that EDT is instrumental in reducing the insurer's risk capital. It is noteworthy that while the silo method tends to yield a more conservative estimate of risk capital, both EDT and copula models lean towards a more aggressive estimation.

Furthermore, when comparing the risk capitals from different models, those derived from EDT are consistently smaller than those from the copula regression models. This is attributable to EDT's ability to capture pairwise dependencies between the two LOBs and the time dependencies of incremental paid losses. Notably, among all models evaluated, DT-CopulaGAN produces the smallest risk capital. This could be due to its use of flexible marginals, such as truncated Gaussians with varying parameters and a Gaussian copula for capturing dependencies between these marginals. Thus, insurers can leverage EDT as an effective tool for risk management, particularly in reducing the required risk capital.
 
Next, we compute the risk capital gain defined in (\ref{eqn:risk_capital_gain}). Note that the risk capital gains for both EDT and copula regression models are compared concerning silo-GLM, which is the industry standard. Table \ref{tab:bootstrap_risk_capital_gain} shows that risk capital gain is crucial when we capture the association between the two LOBs. Further, the larger risk capital gain is obtained for the EDT models compared to the copula regression models. We can associate these gains with the diversification effect in the insurance portfolio. For example, to better take advantage of the diversification effect, the insurer can increase the size of the commercial LOB, which is smaller than the personal LOB.

\begin{table}[H]
\centering
\caption{Risk capital gain for different methods.}
\scalebox{0.85}{
\begin{tabular}{c c c c c c}
\hline
\textbf{Risk capital gain}  & {TVaR (80\%)} & {TVaR (85\%)} &{TVaR (90\%)} &{TVaR (95\%)} & {TVaR(99\%)}\\ 
\hline 
DT-CTGAN vs Silo-GLM & 58.36\% & 58.89\% & 59.64\% & 61.06\% & 61.67\%\\
DT-CopulaGAN vs Silo-GLM& \textbf{61.45\%} & \textbf{61.78\%} & \textbf{62.42\%} & \textbf{64.94\%} & \textbf{66.34\%} \\
DT-Bootstrap vs Silo-GLM & 52.46\% & 52.98\% & 54.99\% & 58.72\% & 63.74\% \\

Product copula vs Silo-GLM & 18.72\% & 19.34\% &20.19\% &22.29\% &25.45\%  \\
Gaussian copula vs Silo-GLM & 22.93\% & 22.98\% & 22.97\% & 24.13\% & 27.21\% \\
Frank copula vs Silo-GLM & 16.31\% & 17.05\% & 18.05\% & 19.59\% & 23.57\% \\
Student's t copula vs Silo-GLM & 17.69\% & 17.52\% & 18.43\% &19.78\% &22.93\% \\

\hline
\end{tabular} 
}
\label{tab:bootstrap_risk_capital_gain}
\end{table}

\section{Simulation Case Study} 

In this section, we further validate our conclusions that the EDT (DT-GAN or DT-bootstrap) produces reduced risk capital through simulation studies. 
We simulate pairs of loss triangles of personal and commercial auto LOBs with ten accident and development years for each simulation run. The details of the simulation setup are provided below.

We begin with the estimated copula regression model for the real data in Section \ref{application}. In this model, we use log-normal and gamma densities for the marginal distributions of standardized incremental paid losses from the personal and commercial LOBs, respectively. To simulate these losses in the loss triangles $\left(Y_{ij}^{(1)}, Y_{ij}^{(2)}\right)$, we first calculate the systematic component ${\eta}_{ij}^{(\ell)} (\ell=1,2)$ from the accident year effect $\alpha_i^{(\ell)}$ (Appendix Table \ref{tab:accident year effect}) and development year effect $\beta_i^{(\ell)}$ (Appendix Table \ref{tab:development year effect}). Next, we simulate $u_{ij}^{(\ell)} (\ell=1,2)$ ($i+j-1 \le I$) from Gaussian copula model $c(\cdot;\theta)$ with dependence parameter $\theta = -0.36$. Then, we transform $u_{ij}^{(\ell)}$ to the upper triangles by inverse function $y_{ij}^{(\ell)}=F^{(\ell)(-1)}(u_{ij}^{(\ell)};{\eta}_{ij}^{(\ell)},{\gamma}^{(\ell)})$, where ${\eta}_{ij}^{(\ell)}=\xi^{(\ell)} + \alpha_i^{(\ell)} + \beta_j^{(\ell)} (\ell=1,2)$. Here, we set the shape parameter $\gamma^{(1)}=0.089$, as estimated in the copula regression model, and larger $\gamma^{(2)}=2$ to account for higher volatility in the commercial LOB. Moreover, ${\eta}_{ij}^{(\ell)}$ are derived from the marginal distribution parameters as follows. $\eta_{ij}=\mu_{ij}$ for a log-normal distribution with location parameter $\mu_{ij}$ and shape parameter $\gamma^{(1)}=\sigma$. For a gamma distribution with location parameter $\mu_{ij}$ and shape parameter $\gamma^{(2)}=\phi$, we use the form $\eta_{ij}=\log(\mu_{ij}\phi)$. Finally, the incremental paid losses, $\left(X_{ij}^{(1)}, X_{ij}^{(2)}\right)$ are obtained by multiplying the simulated $y_{ij}^{(\ell)}$ by the premium for the $i$-th accident year(Appendix Table \ref{tab:premium_table}). Note that in Table Appendix Table \ref{tab:development year effect}, most of the development year effects are negative, which indicates that the incremental paid losses are decreasing with development years. 


Using the above procedure, we simulate the upper and lower parts of each loss triangle. The sum of the lower triangle represents the actual reserve for each loss triangle. We retain only the upper part of all simulated loss triangles to apply the proposed EDT and compare its results with copula regression models. To reflect the multiple companies of real data, we simulate 50 pairs of loss triangles per simulation run.

For each simulation run, we train the DT model using 50 pairs of loss triangles. We use the symmetric loss function for DT in the simulation study because the simulated data was generated using a Gaussian copula, which assumes symmetric dependencies between loss triangles. Since the data does not exhibit inherent asymmetry, applying an asymmetric loss function in this setting would not provide meaningful improvements and could introduce unnecessary distortions. However, in the previous section's real-data analysis, we implemented the asymmetric loss function to capture better potential skewness and tail risks observed in empirical loss triangles. The trained model is then used to predict the reserve for each pair. Additionally, we generate predictive distributions of the total reserve using CTGAN and CopulaGAN for each pair of simulated loss triangles.

\begin{figure}[h!]
    \centering
    \includegraphics[scale=0.6]{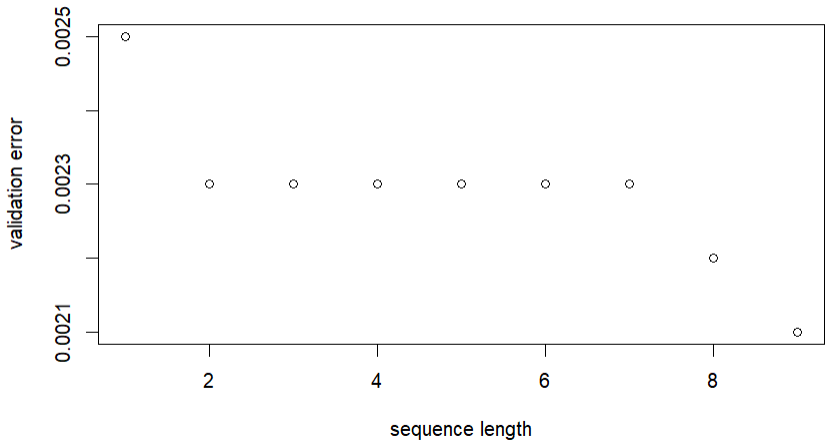}
    \caption{Cross-validation errors for different input sequence lengths in the DT.}
    \label{fig:validation_error_sequence_length}
\end{figure}

\textcolor{black}{Through the simulation study, we examined the impact of input and output sequence lengths on the DT model performance. We generated input and output sequences of varying lengths for each accident year. As shown in Figure \ref{fig:validation_error_sequence_length}, the cross-validation error was evaluated across different sequence lengths, with a length of nine identified as optimal for the DT model.}


Next, we apply the copula regression model separately to each pair of 50 simulated loss triangles. We assume log-normal and gamma distributions for the marginals and evaluate different copula structures, including the product copula, Gaussian copula, Frank copula, and Student's t copula.


In Table \ref{tab:reserve_comparison_simu}, we evaluate the performance of all copula regression and DT models in estimating the total reserve. To compute the true reserve $R^{(1)}$ and $R^{(2)}$, we sum up the expected values of the lower triangle $\exp \left({\mu}_{ij}^{(1)}+\frac{1}{2}\left( {\sigma} \right)^{2} \right)$ and  ${\mu}_{ij}^{(2)} {\phi}$, respectively. 

Here, we compare the predicted reserves with the actual reserves for each pair across all models, including copula regression and DT. In each simulation run, we evaluate 50 pairs of predicted and actual reserves. To quantify the prediction error for each LOB $\ell$ we compute the mean absolute percentage error (MAPE) as defined in \eqref{MAPE}.


\begin{equation}
\label{MAPE}
    \text{MAPE}_{\ell}=\frac{1}{50} \sum_{b=1}^{50} \vert \frac{ \hat{R}_b^{(\ell)}-R^{(\ell)}}{R^{(\ell)}} \vert,
\end{equation}
where $\hat{R}_b^{(\ell)}$ is the predicted reserves for $b^{\text{th}}$ loss triangle from $\ell^{\text{th}}$ LOB and $R^{(\ell)}$ is the true reserve for the $\ell^{\text{th}}$ LOB. Table \ref{tab:reserve_comparison_simu} shows that the DT model outperforms the copula regression models for both LOBs and is particularly effective for the more volatile commercial auto LOB.
The MAPE is almost negligible in personal LOB because the DT has increased flexibility than the copula regression models and captures the time dependence in the sequence input. The MAPE is relatively large for the commercial LOB, which is set to be more volatile than the personal LOB in the simulation setting. 

The performance of the copula regression model can be attributed to its limited flexibility in capturing both the sequence dependence and the pairwise dependence. In particular, the model underestimates the shape parameter of the gamma marginal distribution for the commercial LOB, which plays a crucial role in controlling the dispersion and tail behavior of the distribution. This underestimation leads to significant errors in the predicted reserve, especially for the more volatile commercial LOB.


\begin{table}[h!]
\centering
\begin{threeparttable}
\caption{Performance comparison using the mean absolute percentage error.}
\vspace{0.5cm}
\begin{tabular}{l c c c c c}
\hline
LOB & DT & Product Copula & Gaussian Copula & Frank Copula \\
\hline
Personal Auto & \textbf{0.63\%} & 5.28\% & 5.11\% & 5.07\% \\
Commercial Auto & \textbf{18.67\%} & 27.28\% & 31.85\% & 31.59\% \\
\hline
\end{tabular}
\label{tab:reserve_comparison_simu}

\begin{tablenotes}
      \small
      \item Note:  The best metric for each LOB is in bold. \textcolor{black}{The actual reserves for the personal and commercial LOBs are $\numprint{6 423 246}$ and $\numprint{495 925}$, respectively.}
    \end{tablenotes}
\end{threeparttable}
\end{table}

For each simulated loss triangle pair, we use the DT model's predicted full triangle as input to the GAN models. We then apply CTGAN and CopulaGAN to generate 1,000 synthetic loss triangles per pair and use the DT model to predict reserves for each synthetic triangle.

\begin{figure}[H]
    \centering
    \includegraphics[scale=0.6]{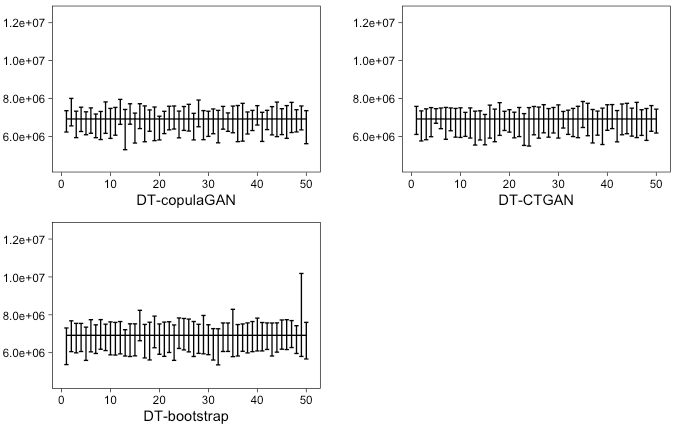}
    \caption{95\% confidence interval for total reserves for different EDT models. \\ {\small Note: The horizontal line indicates the true reserve. The true reserve is within all the 95\% confidence intervals. The average length of confidence intervals are $\numprint{1 413 034}$, $\numprint{1 498 851}$ and $\numprint{1 666 558}$, respectively.} }
    \label{fig:conf_interval_edt50}
\end{figure}

For each of the 50 simulated loss triangle pairs, we construct the predictive distribution of reserves using 1,000 predicted reserves from the corresponding synthetic loss triangles.
Based on the EDT models, the corresponding 95\% confidence intervals for the total reserve are presented in Figure \ref{fig:conf_interval_edt50}, where the horizontal line represents the true reserve of the simulated loss triangles. Notably, we observe that the true reserve falls within all 95\% confidence intervals for all models. Thus, the coverage exceeds the nominal value of 95\%. This over-coverage may be due to the EDT relying on the predicted lower triangle from the DT as input for the GAN models, which could lead to conservative uncertainty estimates, or an insufficient number of synthetic loss triangles, limiting the variability captured in the predictive distribution. Among all EDT models, DT-CopulaGAN produces the narrowest confidence interval. Moreover, comparing GAN and bootstrap, we notice that the block bootstrap with the largest block size has similar properties of confidence intervals of GAN. We expect a smaller block size may lead to the coverage of the interval close to the nominal value. We also expect the coverage of the interval of the GAN to become close to the nominal value when the GAN is modified to accept missing values in the lower triangle.

\begin{figure}[H]
    \centering
    \includegraphics[scale=0.6]{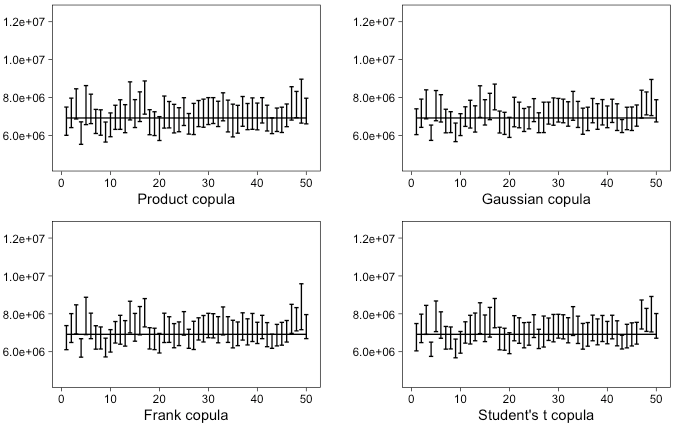}
    \caption{95\% confidence interval for the total reserves for different copula models. \\ {\small Note: The horizontal line indicates the true reserve. The coverage for product copula, Gaussian copula, Frank copula, and Student's t copula are 94\%, 90\%, 88\% ,and 86\%, respectively. The average length of confidence intervals are $\numprint{1 457 826}$, $\numprint{1 291 627}$, $\numprint{1 329 655}$ and $\numprint{1 314 192}$.}}
    \label{fig:conf_interval_copula50}
\end{figure}

For copula regressions, we generate the predictive distribution of the total reserve for each of the 50 simulated loss triangles. For each copula regression model, we conduct 1000 bootstrap simulations to generate the predictive distribution of the total reserve. We present in Figure \ref{fig:conf_interval_copula50} the 95\% confidence interval for the total reserve.
We observe that, for all models, the true reserve falls within most of the 95\% confidence intervals. Among them, the product copula regression model provides the highest coverage, approaching the nominal 95\%, but at the cost of the widest confidence intervals, indicating greater uncertainty. In contrast, the Gaussian copula regression model yields the narrowest confidence intervals, but with a lower coverage rate of 90\%, suggesting it may underestimate reserve variability.

\begin{figure}[H]
    \centering
    \includegraphics[scale=0.46]{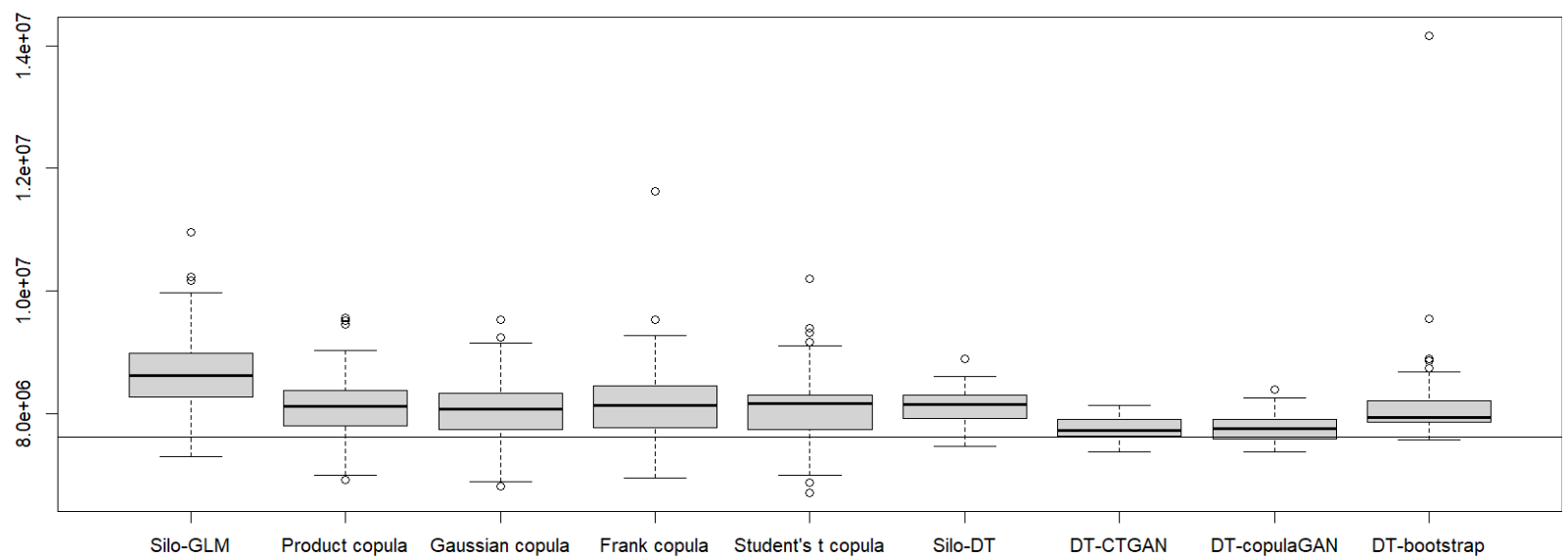}
    \caption{The risk measures at 99\% for different models. The horizontal line indicates the true risk measure.}
    \label{fig:tvar_boxplot_simu50}
\end{figure}

Next, we compute the risk measure for all 50 simulated loss triangle pairs based on their predictive reserve distributions. Figure \ref{fig:tvar_boxplot_simu50} illustrates the box plot of the TVaR measure at the 99\% confidence level for each pair. The boxplots provide a comprehensive comparison across different models: Silo-GLM, Product copula, Gaussian copula, Frank copula, Student's t copula, Silo-DT, DT-CTGAN, DT-copulaGAN, and DT-bootstrap. 

We observe that the median risk measures from all copula regression models are smaller than those from Silo-GLM, while the median risk measures from all EDT models are smaller than those from Silo-DT. This trend can be attributed to the negative association between the two LOBs. Specifically, the copula-based models (Product copula, Gaussian copula, Frank copula, and Student's t copula) demonstrate moderate medians and variability, whereas the DT-based models (DT-CTGAN, DT-copulaGAN, and DT-bootstrap) show the smallest spread, indicating more consistent and lower risk measures.

Furthermore, the spread of the data and the presence of outliers vary across models. Silo-GLM and Silo-DT exhibit larger variability in their risk measures. In contrast, the DT-CTGAN, DT-copulaGAN, and DT-bootstrap models have fewer and lower extreme values, contributing to their overall consistency in risk assessment. For the DT-bootstrap model, we excluded the risk measure from loss triangle 49 in the risk capital calculation due to its outlier status, as indicated in Figure \ref{fig:tvar_boxplot_simu50}. 

Overall, the choice of model significantly influences the assessment of risk measures, with the copula and EDT models providing more conservative and consistent estimates compared to the silo approaches. Between the EDT and copula models, the EDT models generally exhibit lower variability and fewer extreme values, suggesting that EDT models might offer a more robust and reliable assessment of risk in this context.

Additionally, we present the risk capital using the average of the risk measures derived from the 50 simulated loss triangle pairs in Table \ref{tab:risk_capital_simu_50}.  Similar to the real data application, we calculate the risk measures for Silo-GLM and Silo-DT, respectively. Once again, we observe that the risk capital for Silo-DT is smaller than that for Silo-GLM. Furthermore, DT-CopulaGAN consistently generates the smallest risk capital among all the EDT models, aligning with the real data application findings. It's important to note that we do not calculate the risk capital gain in this context because the risk capitals for the silo method vary across the 50 simulated loss triangles. \textcolor{black}{We show the risk capital percentage errors for different models in Table \ref{tab:risk_capital_error_simu_50}. We find that DT-CTGAN and DT-CopulaGAN generate risk capitals that are closest to the true risk capital, particularly in the tail, when compared to all other models.}

\begin{table}[H]
\centering
\caption{Average risk capital from 50 simulated loss triangles.}
\scalebox{1.0}{
\begin{tabular}{c}
\hline
\textbf{Risk capital}  \\
\hline
Silo-DT \\ 
DT-CTGAN \\
DT-CopulaGAN \\ 
DT-Bootstrap \\

Silo-GLM \\
Product copula \\ 
Gaussian copula \\ 
Frank copula \\
Student's t copula \\
\textcolor{black}{True risk capital} \\
\hline 
\end{tabular}%
\begin{tabular}{c c c c c}
\hline
{TVaR (80\%)} & {TVaR (85\%)} &{TVaR (90\%)} &{TVaR (95\%)} & {TVaR(99\%)}\\ 
\hline
  \textbf{198 839} & \textbf{272 500} & \textbf{372 127} & \textbf{531 067} & \textbf{863 805} \\ 
  141 922 & 192 690 & 258 984 & 363 501 & 576 167 \\ 
  140 300 & 190 851 & 257 575 & 361 720 & 569 043 \\ 
  179 339 & 246 619 & 337 177 & 485 248 & 858 038 \\

256 262 & 354 572 & 486 496 & 702 533 & 1 180 878 \\ 
  179 938 & 246 664 & 335 529 & 477 132 & 758 015 \\ 
  158 158 & 217 078 & 295 256 & 422 683 & 700 221 \\ 
  165 128 & 227 321 & 311 099 & 449 677 & 751 712 \\ 
  164 299 & 225 627 & 310 081 & 443 654 & 713 071 \\ 

    106 371 & 146 047 & 199 655 & 282 214 & 467 289 \\

\hline 
\end{tabular}  
}
\label{tab:risk_capital_simu_50}
\end{table}

\begin{table}[ht]
\centering
\caption{Risk capital percentage error for different methods.}
\scalebox{1.0}{
\begin{tabular}{c}
\hline
\textbf{Risk capital}  \\
\hline
Silo-DT \\
DT-CTGAN \\
DT-CopulaGAN \\ 
DT-Bootstrap \\
Silo-GLM \\
Product copula \\ 
Gaussian copula \\ 
Frank copula \\
Student's t copula \\
\hline 
\end{tabular}%
\begin{tabular}{ccccc}
  \hline
{TVaR (80\%)} & {TVaR (85\%)} &{TVaR (90\%)} &{TVaR (95\%)} & {TVaR(99\%)}\\ 
  \hline
    86.93\% & 86.58\% & 86.38\% & 88.18\% & 84.85\% \\ 
  33.42\% & 31.94\% & 29.72\% & 28.80\% & 23.30\% \\ 
  \textbf{31.90\%} & \textbf{30.68\%} & \textbf{29.01\%} & \textbf{28.17\%} & \textbf{21.78\%} \\ 
  76.50\% & 77.47\% & 78.52\% & 84.02\% & 103.63\% \\
140.91\% & 142.78\% & 143.67\% & 148.94\% & 152.71\% \\ 
  69.16\% & 68.89\% & 68.05\% & 69.07\% & 62.22\% \\ 
  48.68\% & 48.64\% & 47.88\% & 49.77\% & 49.85\% \\ 
  55.24\% & 55.65\% & 55.82\% & 59.34\% & 60.87\% \\ 
  54.46\% & 54.49\% & 55.31\% & 57.20\% & 52.60\% \\ 

\hline
\end{tabular}
}
\label{tab:risk_capital_error_simu_50}
\end{table}

\section{Summary and Discussion}



The proposed Extended Deep Triangle (EDT) method enhances the Deep Triangle model by effectively capturing interdependencies between loss triangles from different LOBs, leveraging diversification effects within an insurance portfolio. \textcolor{black}{In the EDT framework, incremental paid losses from both LOBs are used as training inputs, with the model designed to minimize asymmetric prediction errors}.

The DT model estimates reserves for both LOBs, while the EDT models—DT-CTGAN and DT-CopulaGAN generate predictive distributions for the total reserve. We demonstrate that DT requires sequence-based input, whereas GAN-based models operate on tabular data, ensuring both approaches effectively capture dependencies and improve reserve estimation. Additionally, block bootstrapping of the training data is used with the DT model to generate the predictive distribution of the reserve (DT-bootstrap). 


\textcolor{black}{A crucial aspect of this paper is assessing the diversification benefits of EDT in multivariate loss reserving and risk capital analysis. This is achieved by comparing risk measures and risk capital derived from the "silo" method against those obtained from EDT or copula regression approaches. This comparison highlights the potential advantages of employing EDT, a more interconnected and sophisticated modeling technique, in managing insurance portfolio risks.
  }
  

To evaluate the practical effectiveness of EDT, we apply it to loss triangles from 30 companies in the NAIC database. For comparison, we also include copula regression in our study. EDT outperforms alternative models by yielding the smallest bias between predicted and true reserves for both LOBs. Moreover, both EDT variants—DT-GAN and DT-Bootstrap—consistently produce lower risk capital estimates than industry standards. These results highlight the potential benefits of integrating advanced modeling techniques for more accurate reserve estimation and enhanced risk management in the insurance industry.
 
Through the real data applications and simulations study, we discerned certain limitations of copula regression. One reason for the relatively large bias in copula regression is using a single pair of loss triangles. Future extensions may involve seemingly unrelated regressions and mixed models to address the heterogeneity in data from multiple companies \citep{zellner1962efficient}. Another notable constraint is that complex dependencies may not be captured in copula regression attributed to the Fr{\'e}chet-Hoeffding bounds on the dependence parameter \citep{schweizer2011probabilistic}. Additionally, the potential over-parameterization of the copula regression model, particularly given the limited pairs of observations available in the LOBs, poses a challenge. To mitigate this, regularization techniques such as the least absolute shrinkage and selection operator (LASSO) can be applied. By imposing an $L_1$ norm constraint on the parameters within the loss function, LASSO facilitates parameter shrinkage, effectively reducing overfitting \citep{tibshirani1996regression}. These regularization methods, akin to those used in generalized linear models \citep{Taylor2019loss}, play a crucial role in improving the robustness of copula regression models, especially when working with limited data. 




In the context of EDT, it is valuable to delve into the dependence structure, strength, and dependence sign captured between loss triangles. Future extensions may incorporate mutual information to calculate expected and point-wise dependencies \citep{belghazi2018mutual, tsai2020neural}, along with the appropriate statistical measures for structure and signs, to enhance the understanding of the interplay between loss triangles in risk capital analysis.

In summary, the EDT framework shows strength and potential in multivariate loss reserving and risk capital analysis by providing a smaller bias in reserve prediction and larger diversification benefits. This flexible framework allows its application in various settings in actuarial science, such as rate-making and reinsurance. This adaptability showcases the EDT versatility across diverse insurance scenarios by complementing the DT model.

\section*{Acknowledgements}

This work was funded by the Start-up funds, the Faculty of Science at McMaster University (Pratheepa Jeganathan), and the Natural Sciences and Engineering Research Council of Canada (Anas Abdallah). Support provided by SHARCNET (sharcnet.ca) and the Digital Research Alliance of Canada (alliancecan.ca) partly enabled this research.

\section*{Authors' contribution}

Conceptualization, A.A. and P.J.; methodology, P.C., A.A., and P.J.; software, P.C.; validation, P.C.; formal analysis, P.C.; investigation, P.C.; resources, A.A. and P.J.; writing—original draft preparation, P.C.; writing—review and editing, A.A., P.J. and P.C.; supervision, A.A. and P.J. All authors have read and agreed to the manuscript.

\clearpage

\bibliography{manuscript}

\clearpage

\begin{appendices}
\section{Datasets}

We present the incremental paid losses dataset from a major US property-casualty insurer. The dataset includes loss triangles from two lines of business (LOBs): personal and commercial LOBs.

\label{app:data}
\begin{table}[h!]
\caption{Incremental paid losses ($X_{ij}^{(1)}$) for personal auto LOB.}
\small
\label{tab:data_shi_lob1}
\scalebox{0.9}{
\begin{tabular}{rrrrrrrrrrrr}
\hline \hline
year & premium & 1    & 2    & 3   & 4   & 5   & 6   & 7  & 8  & 9  & 10 \\
\hline 
1988 & 4 711 333 & 1 376 384 & 1 211 168 & 535 883 & 313 790 & 168 142 & 79 972 & 39 235 & 15 030 & 10 865 & 4 086 \\ 
  1989 & 5 335 525 & 1 576 278 & 1 437 150 & 652 445 & 342 694 & 188 799 & 76 956 & 35 042 & 17 089 & 12 507 &  \\ 
  1990 & 5 947 504 & 1 763 277 & 1 540 231 & 678 959 & 364 199 & 177 108 & 78 169 & 47 391 & 25 288 &  &  \\ 
  1991 & 6 354 197 & 1 779 698 & 1 498 531 & 661 401 & 321 434 & 162 578 & 84 581 & 53 449 &  &  &  \\ 
  1992 & 6 738 172 & 1 843 224 & 1 573 604 & 613 095 & 299 473 & 176 842 & 106 296 &  &  &  &  \\ 
  1993 & 7 079 444 & 1 962 385 & 1 520 298 & 581 932 & 347 434 & 238 375 &  &  &  &  &  \\ 
  1994 & 7 254 832 & 2 033 371 & 1 430 541 & 633 500 & 432 257 &  &  &  &  &  &  \\ 
  1995 & 7 739 379 & 2 072 061 & 1 458 541 & 727 098 &  &  &  &  &  &  &  \\ 
  1996 & 8 154 065 & 2 210 754 & 1 517 501 &  &  &  &  &  &  &  &  \\ 
  1997 & 8 435 918 & 2 206 886 &  &  &  &  &  &  &  &  &  \\ 
\hline    
\end{tabular}
}
\end{table}

\begin{table}[h!]
\caption{Incremental paid losses ($X_{ij}^{(2)}$) for commercial auto LOB. }
\small
\label{tab:data_shi_lob2}
\scalebox{0.9}{
\begin{tabular}{rrrrrrrrrrrr}
\hline \hline
year & premium & 1    & 2    & 3   & 4   & 5   & 6   & 7  & 8  & 9  & 10 \\
\hline
1988 & 267 666 & 33 810 & 45 318 & 46 549 & 35 206 & 23 360 & 12 502 & 6 602 & 3 373 & 2 373 & 778 \\ 
  1989 & 274 526 & 37 663 & 51 771 & 40 998 & 29 496 & 12 669 & 11 204 & 5 785 & 4 220 & 1 910 &  \\ 
  1990 & 268 161 & 40 630 & 56 318 & 56 182 & 32 473 & 15 828 & 8 409 & 7 120 & 1 125 &  &  \\ 
  1991 & 276 821 & 40 475 & 49 697 & 39 313 & 24 044 & 13 156 & 12 595 & 2 908 &  &  &  \\ 
  1992 & 270 214 & 37 127 & 50 983 & 34 154 & 25 455 & 19 421 & 5 728 &  &  &  &  \\ 
  1993 & 280 568 & 41 125 & 53 302 & 40 289 & 39 912 & 6 650 &  &  &  &  &  \\ 
  1994 & 344 915 & 57 515 & 67 881 & 86 734 & 18 109 &  &  &  &  &  &  \\ 
  1995 & 371 139 & 61 553 & 132 208 & 20 923 &  &  &  &  &  &  &  \\ 
  1996 & 323 753 & 112 103 & 33 250 &  &  &  &  &  &  &  &  \\ 
  1997 & 221 448 & 37 554 &  &  &  &  &  &  &  &  &  \\
  \hline
\end{tabular}
}
\end{table}

\section{Copula Regression and Bootstrapping}
\label{section:copula_regression}
\subsection{Copula Model}
Next, we briefly describe the copula regression and parametric bootstrapping methods used for comparison with the proposed EDT method.

Now we consider the cumulative distribution function (CDF) of $Y_{ij}$, which is given by
\begin{equation}
F_{ij}=\textrm{Prob} (Y_{ij}\le y_{ij})=
F(y_{ij};\eta_{ij},\gamma),
\end{equation}
where $\eta_{ij}$ denotes the systematic component, which determines the location and $\gamma$ determines the shape. 

We assume $\gamma$ is the same for all the cells $(i,j)$ for each loss triangle. Now we model the systematic component $\eta_{ij}$ using  $\alpha_i (i \in {1,2,...,10})$ and $\beta_j (j \in {1,2,...,10})$ as predictors that characterize the effect of the accident year and the development year corresponding to $Y_{ij}$ as in \eqref{eq:eta}. 

\begin{equation}
\label{eq:eta}
\eta_{ij}=\xi + \alpha_i + \beta_j,  
\end{equation}
where $\xi$ is the intercept and $\alpha_1=0$ and $\beta_1=0$ are constraints for parameter identification.

 We use the goodness-of-fit test to choose the distribution for $Y_{i,j}^{(1)}$. It identifies that  $Y_{i,j}^{(1)}$ and $Y_{i,j}^{(2)}$ follow log-normal and gamma distributions, respectively. Let's consider the probability density function (PDF) of the log-normal distribution for $Y_{ij}^{(1)}$ 

\begin{equation}
    f_{i j}^{(1)}\left(y_{i j}^{(1)}\right)=\frac{1}{y_{i j}^{(1)} \sigma \sqrt{2 \pi}} e^{-\frac{1}{2}
    \left(\frac{\log\left(y_{ij}^{(1)}\right)-\mu_{ij}^{(1)}}{\sigma}\right)^{2}}, \quad y_{i j}^{(1)}>0,
\end{equation}
where  $\mu_{ij}^{(1)}$ is the location and  $\sigma > 0$ is the shape. Thus, the systematic component is $\eta_{ij}^{(1)}=\mu_{ij}^{(1)}$.

Next, the gamma PDF for $Y_{ij}^{(2)}$ is given by

\begin{equation}
    f_{i j}^{(2)}\left(y_{i j}^{(2)}\right)=\left(\frac{y_{i j}^{(2)}}{\mu_{i j}^{(2)}}\right)^{\phi} \frac{e^{-\frac{y_{i j}^{(2)}}{\mu_{i j}^{(2)}}}}{\Gamma\left(\phi\right) y_{i j}^{(2)}}, \quad y_{i j}^{(2)} > 0,
\end{equation}
where $\phi >0$ is the shape and  $\mu_{ij}^{(2)} > 0$ is the location. Thus, the systematic component is $\eta_{ij}^{(2)}=\log\left(\mu_{ij}^{(2)}\phi\right)$ to ensure positive predicted values \citep{Abdallah2015}.

In addition to the specified marginal densities, we assume that $Y_{i j}^{(1)}$ and $Y_{i j}^{(2)}$ from different LOBs with the same accident and development year are dependent. This is called pair-wise dependence. Moreover, we consider the copulas to model the dependence structure between the two lines of business \citep{Shi2011}. 
 
Next, we write the joint distribution of $\left( Y_{ij}^{(1)},Y_{ij}^{(2)}\right)$ using copulas based on Sklar's theorem \citep{Nelsen_copula_2006} as follows

\begin{equation}
F_{i j}\left(y_{i j}^{(1)}, y_{i j}^{\left(2\right)}\right)=\operatorname{Prob}\left(Y_{i j}^{(1)} \leq y_{i j}^{(1)}, Y_{i j}^{\left(2\right)} \leq y_{i j}^{\left(2\right)}\right)=C\left(F_{i j}^{(1)}(y_{i j}^{(1)}), F_{i j}^{\left(2\right)}(y_{i j}^{(2)}) ; {\theta}\right),
\label{eqn:joint_distribution}
\end{equation}
where $F_{ij}^{(1)}$ and $F_{ij}^{(2)}$ are the marginal distributions for $Y_{i j}^{(1)}$ and $Y_{i j}^{(2)}$, respectively, and $C(\cdot, {\theta})$ is the copula function such that $C(\cdot, {\theta}): [0,1]^2 \mapsto [0,1]$ with parameter ${\theta}$.

By getting derivative of \eqref{eqn:joint_distribution} with respect to $Y_{i j}^{(1)}$ and $Y_{i j}^{(2)}$, we get the join PDF for $\left( Y_{ij}^{(1)},Y_{ij}^{(2)}\right)$ in \eqref{eqn:joint_PDF}.

\begin{equation}
\label{eqn:joint_PDF}
f_{i j}\left(y_{i j}^{(1)}, y_{i j}^{\left(2\right)}\right)=c\left(F_{i j}^{(1)}, F_{i j}^{\left(2\right)} ; {\theta}\right) \prod_{\ell=1}^{2} f_{i j}^{(\ell)},
\end{equation}
where $c(\cdot)$ denotes the PDF corresponding to copula $C(\cdot)$ and $f_{i j}^{(\ell)}$ denotes the PDF associated with the marginal distribution $F_{i j}^{(\ell)}$.

Next, we use the maximum likelihood method to estimate the parameters in the regression model in \eqref{eq:eta} using the copula density in \eqref{eqn:joint_PDF}. We denote estimators as $\hat{\mu}_{ij}^{(\ell)}$, $\hat{\sigma}$ and $\hat{\phi}$.

The log-likelihood for all joint $\left(Y_{i j}^{(1)}, Y_{i j}^{(2)}\right)$ is given by

\begin{align}
\begin{split}
L(\eta_{i j}^{(1)},\gamma^{(1)},\eta_{i j}^{(2)},\gamma^{(2)},\theta)&=\sum_{i=1}^{I} \sum_{j=1}^{I+1-i}\log \left(c\left(F_{i j}^{(1)}, F_{i j}^{\left(2\right)} ; {\theta}\right)\right) \\ &+\sum_{i=1}^{I} \sum_{j=1}^{I+1-i} \sum_{\ell=1}^{2} \log \left(f_{i j}^{(\ell)}\right),
\end{split}
\end{align}
where $\gamma^{(1)} = \sigma$, $\gamma^{(2)} = \phi$, $\eta_{ij}^{(1)}=\mu_{ij}^{(1)}$, $\eta_{ij}^{(2)}=\log\left(\mu_{ij}^{(2)}\phi\right)$, and $\eta_{ij}$ is a function of $\alpha_{i}$ and $\beta_{j}$ as in regression model \eqref{eq:eta}.

For the log-normal distribution, the $Y_{ij}^{(1)}$ is estimated by $\hat{Y}_{ij}^{(1)} = \exp \left( \hat{\mu}_{ij}^{(1)}+\frac{1}{2}\left( \hat{\sigma} \right)^{2} \right)$ and for the gamma distribution, $\hat{Y}_{ij}^{(2)}=\hat{\mu}_{ij}^{(2)} \hat{\phi} $. To estimate $X_{i j}^{(\ell)}$, lower triangle, we multiply  $\hat{Y}_{ij}^{(\ell)}$ by the corresponding premium $\omega_{i}^{(\ell)}$.

\subsection{Predictive Distribution of the Total Reserve}\label{risk_measures}

The two most popular approaches to generating the predictive distribution of the reserve based on the copulas are simulation and parametric bootstrapping.

Simulation is based on the estimated copula regression model, in which we use the Monte Carlo simulation to generate the predictive distribution of the reserve. The simulation is summarized as the following procedure \citep{Shi2011}:
\begin{itemize}
\item[$(1)$] Simulate $\left(u_{ij}^{(1)},u_{ij}^{(2)} \right)$ ($i+j-1>I$) from estimated copula function $C(\cdot;\hat{\theta})$.
\item[$(2)$] Transform $u_{ij}^{(\ell)}$ to predictions of the lower triangles by inverse function $y_{ij}^{*(\ell)}=F^{(\ell)(-1)}(u_{ij}^{(\ell)};\hat{\eta}_{ij}^{(\ell)},\hat{\gamma}^{(\ell)})$, where $\hat{\eta}_{ij}^{(\ell)}=\hat{\xi}^{(\ell)} + \hat{\alpha}_i^{(\ell)} + \hat{\beta}_j^{(\ell)}$ .
\item[$(3)$] Obtain a prediction of the total reserve by
\begin{equation*}
\sum_{\ell=1}^{2} \sum_{i=2}^{I} \sum_{j=I-i+2}^{I} \omega_i^{(\ell)} y_{ij}^{*(\ell)}.
\end{equation*}
\end{itemize}

Repeat Steps (1) - (3) many times to obtain the predictive distribution of $R$. However, the limitation of Monte Carlo simulation is the inability to incorporate estimated parameter uncertainty. To address this constraint, we consider the parametric bootstrapping.

In the parametric bootstrapping, we generate a new upper triangle for each simulation with estimated parameters and fit the corresponding copula regression model to this new upper triangle \citep{taylor2007synchronous, Shi2011}. 
The detailed algorithm is as follows:

\begin{itemize}
\item[$(1)$] Simulate $\left(u_{ij}^{(1)},u_{ij}^{(2)} \right)$ ($i+j-1\le I$) from estimated copula function $C(\cdot;\hat{\theta})$.
\item[$(2)$] Transform $u_{ij}^{(\ell)}$ to estimate the upper triangles by inverse transform $y_{ij}^{*(\ell)}=F^{(\ell)(-1)}(u_{ij}^{(\ell)};\hat{\eta}_{ij}^{(\ell)},\hat{\gamma}^{(\ell)})$, where $\hat{\eta}_{ij}^{(\ell)}=\hat{\xi}^{(\ell)} + \hat{\alpha}_i^{(\ell)} + \hat{\beta}_j^{(\ell)}$ .
\item[$(3)$] Generate an estimate of the total reserve using $y_{ij}^{*(\ell)}$ from step (2).
\begin{itemize}
    \item Estimate the parameters $\hat{\eta}_{ij}^{*(\ell)}$,$\hat{\gamma}^{*(\ell)}$ and $\hat{\theta}^*$ by performing MLE for the copula regression model for $y_{ij}^{*(\ell)}$.
    \item Use $\hat{\eta}_{ij}^{*(\ell)}$,$\hat{\gamma}^{*(\ell)}$ and $\hat{\theta}^*$ to simulate the lower triangle, $y_{ij}^{**(\ell)}$ using the simulation Steps (1) and (2).
    \item Obtain a prediction of the total reserve by
    \begin{equation*}
    \sum_{\ell=1}^{2} \sum_{i=2}^{I} \sum_{j=I-i+2}^{I} \omega_i^{(\ell)} y_{ij}^{**(\ell)}.
    \end{equation*}
\end{itemize}
\end{itemize}

Repeat Steps (1)-(3) many times to obtain a predictive reserve distribution.



\section{Dependence analysis between the personal and commercial auto LOBs}
\label{app:kendall_tau}
We compute Kendall's tau on the residuals from the log-normal and gamma regression models for the standardized incremental paid losses from personal ($Y_{ij}^{(1)}$)  and commercial ($Y_{ij}^{(2)}$) auto LOBs. Note that the analysis is performed on the residuals because we want to take out the effects of covariates (accident year and development year effects). For the log-normal regression, the residual is $\hat{\epsilon}_{ij}^{(1)}=(\text{ln} y_{ij}^{(1)}-\hat{\mu}_{ij}^{(1)})/\hat{\sigma}$, and the residual is $\hat{\epsilon}_{ij}^{(2)}=y_{ij}^{(2)}/\hat{\mu}_{ij}^{(2)} $ for the gamma regression. The computed Kendall's tau is -0.1562, suggesting a negative association between personal and commercial LOBs. 

\section{Summary statistics for copula regression}

\begin{table}[h!]
\centering
\caption{Summary statistics for different fitted copula regression models.}
\begin{tabular}{l c c c c}
\hline 
 & \multicolumn{4}{c}{Copula} \\ 
\cline{2-5}
 & Product & Gaussian & Frank & Student's t \\ 
\hline 
Dependence Parameter ($\theta$) & . & -0.3656 & -2.7977 & -0.2657 \\ 
Log-Likelihood &346.6  &350.4  &350.3  & 355.4\\
AIC &-613.2  &-618.9  &-618.5 &-626.9 \\
BIC & -505.2 & -508.2 & -507.8 & -513.5 \\
\hline 
\end{tabular}  
\label{tab:copula_fit_statistics}
\end{table}

\section{Inference on the copula parameter}
\label{app:dependence_theta}
We present the 95\% confidence intervals for the copula parameter from the copula regression models in Table \ref{app_tab:theta_confidence_interval}.

\begin{table}[h!]
\centering

\begin{threeparttable}
\caption{{95\%} confidence intervals for copula parameter.}
\label{app_tab:theta_confidence_interval}
\begin{tabular}{l  c c}
\hline
 & 2.5\% percentile & 97.5 \% percentile \\
\hline
Gaussian copula & -0.605 & -0.136 \\
Frank copula & -4.99 & -1.38  \\
Student's t copula & -0.64 & 0.27  \\
\hline
\end{tabular}

\begin{tablenotes}
      \small
      \item Note: The Student's t copula parameter confidence interval includes 0.
    \end{tablenotes}
\end{threeparttable}

\end{table}

\section{Copula regression using loss triangles from 30 companies}
\label{app:copula_regression30}
Here we consider modeling the systematic component $\eta_{ijc}$ using accident year effect  $\alpha_i (i \in {1,2,...,10})$, development year effect $\beta_j (j \in {1,2,...,10})$, and company effect $b_c (c \in {1,2,..,30})$ as in \eqref{eq:eta_30}. 

\begin{equation}
\label{eq:eta_30}
\eta_{ijc}=\xi + \alpha_i + \beta_j+b_c,  
\end{equation}
where $b_c$ is an additional predictor that characterizes the company effect.

We use Gaussian copula to capture the dependence between the two LOBs, and the estimated reserves are $\numprint{6 823 325}$ and $\numprint{370 386}$, respectively. The percentage errors of actual and estimated reserves for the two LOBs are $-15.62\%$ and $16.33\%$, respectively.

\section{Simulation setting}

We present the true values of the parameters used in the simulation study.

\begin{table}[H]
\centering
\caption{Accident year effect $\alpha_i$}
\label{tab:accident year effect}
\begin{tabular}{ccc}
\hline \hline
        & personal auto & commercial auto \\ \hline 
year 2  & -0.03         & -0.14           \\ 
year 3  & -0.03         & -0.15           \\ 
year 4  & -0.13         & -0.30           \\ 
year 5  & -0.17         & -0.29           \\ 
year 6  & -0.18         & -0.27           \\ 
year 7  & -0.18         & -0.14           \\ 
year 8  & -0.24         & -0.10           \\ 
year 9  & -0.27         & 0.17            \\ 
year 10 & -0.21         & -0.12           \\ 
\hline
\end{tabular}
\end{table}

\begin{table}[H]
\centering
\caption{Development year effect $\beta_j$}
\label{tab:development year effect}
\begin{tabular}{ccc}
\hline \hline
       & personal auto & commercial auto \\ \hline
dev 2  & -0.23         & 0.20            \\ 
dev 3  & -1.05         & -0.02           \\ 
dev 4  & -1.65         & -0.41           \\ 
dev 5  & -2.26         & -1.06           \\ 
dev 6  & -3.02         & -1.47           \\ 
dev 7  & -3.68         & -2.10           \\ 
dev 8  & -4.50         & -2.81           \\ 
dev 9  & -4.91         & -3.12           \\ 
dev 10 & -5.92         & -4.18           \\ 
\hline 
\end{tabular}
\end{table}

\begin{table}[H]
\centering
\caption{Premium $\omega_i$}
\label{tab:premium_table}
\begin{tabular}{ccc}
\hline \hline 
        & personal auto & commercial auto \\ \hline
year 1  & 4 711 333     & 267 666         \\ 
year 2  & 5 335 525     & 274 526         \\ 
year 3  & 5 947 504     & 268 161         \\ 
year 4  & 6 354 197     & 276 821         \\ 
year 5  & 6 738 172     & 270 214         \\ 
year 6  & 7 079 444     & 280 568         \\ 
year 7  & 7 254 832     & 344 915         \\ 
year 8  & 7 739 379     & 371 139         \\ 
year 9  & 8 154 065     & 323 753         \\ 
year 10 & 8 435 918     & 221 448         \\ \hline  
\end{tabular}
\end{table}

\end{appendices}

\end{document}